\documentclass{article}

\usepackage{arxiv}

\usepackage[utf8]{inputenc} 
\usepackage[T1]{fontenc}    
\usepackage[colorlinks=true, linkcolor=blue, citecolor=blue, urlcolor=blue, pdfborder={0 0 0}]{hyperref}
\usepackage{url}            
\usepackage{booktabs}       
\usepackage{amsfonts}       
\usepackage{nicefrac}       
\usepackage{microtype}      
\usepackage{lipsum}		
\usepackage{graphicx}
\usepackage[numbers,sort&compress]{natbib}
\usepackage{doi}
\usepackage{float}
\usepackage{amsthm}
\usepackage{physics}
\usepackage{bm}
\usepackage{arydshln}
\usepackage{caption}

\providecommand{\JournalTitle}[1]{#1}

\theoremstyle{definition}

\newcommand{\supplementaryfigures}{
  \renewcommand{\thefigure}{S\arabic{figure}}
  \setcounter{figure}{0}
}
\newcommand{\supplementarytables}{
  \renewcommand{\thetable}{S\arabic{table}}
  \setcounter{table}{0}
}





\title{Collective evolutionary dynamical-systems games: Emergent turn-taking under anonymous monitoring}

\author{Kenji Itao\\
    Frontier Research Institute for Interdisciplinary Sciences,\\
    Tohoku University, Aramaki aza Aoba 6-3, Aoba-ku, Sendai 980-8578, Japan\\
	Computational Group Dynamics Collaboration Unit,\\
    RIKEN Center for Brain Science, Japan.\\
	\texttt{kenji.itao@tohoku.ac.jp} \\
	\And
	Kunihiko Kaneko\\
	The Niels Bohr Institute, \\
    University of Copenhagen, \\
    Jagtvej 128, Copenhagen, 2200-DK, Denmark.\\
	\texttt{kunihiko.kaneko@nbi.ku.dk} \\
}

\renewcommand{\shorttitle}{Collective evolutionary dynamical-systems games}

\hypersetup{
pdftitle={Collective evolutionary dynamical-systems games},
pdfauthor={Kenji Itao, Kunihiko Kaneko},
pdfkeywords={social institutions, evolutionary game theory, dynamical-systems game, statistical physics},
}

\begin{document}
\maketitle

\begin{abstract}
	Sustainable resource use in large societies requires institutions that define and enforce acceptable behaviour while keeping monitoring costs manageable. Here we show that such institutions can arise even when users observe only anonymised group-level information. Using the \emph{evolutionary dynamical-systems game} framework, we model a common-pool resource management game. In the model, each player's harvesting decisions depend on the resource abundance, the player's wealth, and the group-average wealth. Strategies are encoded as two‑parameter decision-making functions that mutate across generations. Evolutionary simulations reveal that players self-organise into clusters that alternate harvesting turns. A simple rule emerges, \emph{wait when wealthy, harvest when below average,} which generates turn-taking patterns and robustness against selfish overharvesters. While the majority cluster tends to exploit the minority, moderate diversity in strategies allows ``turn‑taking of turns'' between the majority and minority roles, improving efficiency, equity, and robustness. Moderate stochasticity in both resource dynamics and peer monitoring can enhance efficiency. The conditions for sustaining these institutions depend strongly on group size and demographic history. Gradual group expansion allows sustainability at substantially larger group sizes. These results provide a mechanistic account of the self-organisation of institutions under anonymous monitoring.
\end{abstract}


\section*{Introduction}
Coordination in large societies requires social institutions that define and enforce acceptable behaviour \cite{greif1998historical, north1981structure, north2005understanding, powers2016institutions}. Although selfish actions may maximise individual payoffs, institutions realign incentives toward collective welfare, including sustainable resource use. Even without private property rights or top-down governmental regulation, many indigenous societies have maintained sustainability through self-organised institutions \cite{ostrom1990governing, ostrom1994rules, haller2002common, garnett2018spatial}.
Institutions for resource management are often forged through long-term trial and error, as communities calibrate the frequency of resource use and develop sanctioning systems that enforce compliance. Hence, codification typically comes only \emph{ex post} \cite{rivera1998water,lobe2004padu}. Yet two questions remain unsolved: how such institutions emerge from individual interactions and how the difficulty of maintenance scales with population size \cite{ostrom1990governing}.

Effective governance of common-pool resources, such as forests, fisheries, and irrigation systems, requires social institutions to prevent the tragedy of the commons, in which individually rational resource use leads to collectively undesirable overuse \cite{hardin1968tragedy, ostrom1990governing, cox2010review}. Institutions frequently stipulate temporal windows, including seasonal closures, during which resource use is either permitted or prohibited. For example, some communities allow fishing only in even-numbered years \cite{bayliss2010managing}, restrict farming to designated weekdays \cite{bayliss2010managing, osei2017taboos}, or enforce turn-taking in water distribution \cite{ostrom1990governing, lobe2004padu}. The acceptability of actions also varies across contexts. During periods of abundance, water allocations are proportional to land area, whereas during scarcity, a minimum per-capita share is guaranteed \cite{guillet1992covering}. Therefore, cooperative behaviour in such settings is inherently context-dependent and cannot be reduced to a fixed cooperate–defect dichotomy \cite{ostrom1990governing, cox2010review}.

Enforcing such rules requires monitoring, whose informational demands increase with group size \cite{ostrom1990governing, gardner1990nature, agrawal2001group}. If \(M\) users individually monitor one another, the total cost scales as \(\mathcal{O}(M^2)\). Even with centralised record keeping, collecting behavioural records costs at least $\mathcal{O}(M)$, with further costs incurred to deter corruption \cite{ostrom1990governing}. In many field settings, however, identifying specific overusers is difficult or prohibitively costly, whereas communities can observe aggregate indicators, such as groundwater levels, tree felling, or total catches, at a cost of only $\mathcal{O}(1)$ \cite{reddy2014groundwater, eisenbarth2021can, mbaiwa2014community, pons2022trade, farmer2020forecasting}. Although anonymous or imperfect monitoring has been extensively studied in repeated games with static payoffs \cite{fudenberg2011folk, bernard2016folk, goldlucke2018discounted}, little is known about its role in dynamically renewing resource systems. This raises the question of whether common-pool resource governance can rely solely on coarse-grained signals.

A trade-off exists between monitoring simplicity and resource-use efficiency. For instance, closed seasons can simplify enforcement because any harvesting during the closure is a clear violation.

However, field observations report that harvesting is intensified before and after the closure \cite{zhang2022dynamics}, and daily harvesting effort during openings was four to 60 times that on continuously open sites \cite{cohen2013fishing}. Biomass dynamics models predict that pulsed intensive harvesting reduces efficiency compared to continuous harvesting \cite{colvin2012semidiscrete}.
By contrast, turn-taking among user clusters distributes harvesting more evenly over time, thereby mitigating such pulse effects, but it also makes violations harder to detect. We therefore investigate whether such turn-taking arises and endures through spontaneous cluster formation when only anonymous information is available.

Additionally, the impact of heterogeneity and noise on collective action is contested. Sociocultural diversity is argued to impede resource governance, whereas moderate cognitive or cultural variation can enhance collective performance \cite{poteete2004heterogeneity, gehrig2019sociocultural, baggio2019importance, van2021heterogeneity}.
Noise in monitoring could hinder collective action by obscuring individual responsibility and by making enforcement contingent on imperfect signals \cite{fudenberg2011folk, bernard2016folk, goldlucke2018discounted}.
Here, we examine how heterogeneity and noise affect sustainability.

To address these questions, we model common-pool resource management as an evolutionary game with environmental feedback. Current payoffs depend on resource abundance shaped by past actions, and criteria of cooperation are context dependent \cite{ostrom1990governing, cox2010review, tilman2020evolutionary, farahbakhsh2022modelling}. Conventional games with fixed cooperation--defection dichotomy and static payoff cannot capture this structure. We therefore use evolutionary dynamical-systems game theory \cite{akiyama2000dynamical, akiyama2002dynamical, itao2025self}, in which players' context-dependent decision rules evolve, and their actions reshape future payoffs by altering the environment. Earlier work on the two-player case showed that such evolution can generate institutions with emergent criteria of cooperation and sanctions against violators \cite{itao2025self}. The collective setting considered here is qualitatively different. Once \(M \ge 3\), the group average no longer reveals individuals' behaviour, so anonymity becomes a genuine informational constraint.

We use a multilevel evolutionary framework in which resource use and social interactions occur within groups, whereas individuals (but not groups) reproduce. Differences in group mean fitness can thus favour decision rules that restrain within-group overuse. While group membership may persist across generations \cite{takeuchi2017origin, zadorin2023multilevel, itao2020evolution, itao2024formation} or be randomly reassigned in each generation \cite{nowak2010evolution}, persistent membership strengthens within-group correlations and can promote group-beneficial traits. Here, we use random reassignment as a conservative setting for the evolution of sustainable resource use.
In this framework, the number of groups and the group size determine the relative strengths of intergroup and intragroup selection, respectively, and the scaling between them quantifies the conditions for structure formation \cite{takeuchi2017origin, takeuchi2022scaling, traulsen2006evolution, itao2020evolution, itao2021evolution, itao2022emergence, itao2024formation}. We therefore measure the minimum number of groups required for sustainability to quantify how the difficulty of maintaining institutions scales with group size.

In this study, we derive four key principles:
(i) Anonymous monitoring of group-average wealth can support sustainability. (ii) Institutions based on the simple rule \emph{wait when wealthy, harvest when below average} emerge spontaneously and generate self-organised turn-taking and robustness against cheaters. (iii) Strategic diversity and noise expand the repertoire of turn-taking patterns and can improve efficiency, equity, and robustness to invasion. (iv) The minimum number of groups required for sustainability scales exponentially but gradual expansion substantially slows this increase, highlighting strong path dependence.

\begin{figure}[tb]
  \centering
   \includegraphics[width=.7\linewidth]{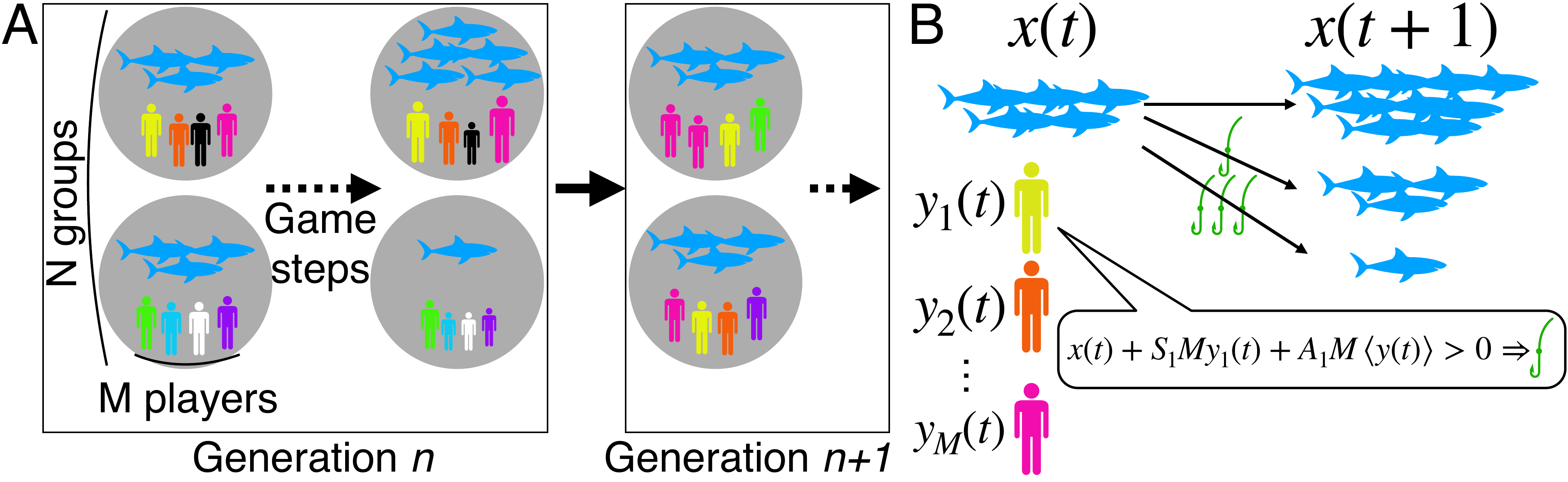}
   \caption{Schematic of the model. (A) Evolutionary dynamical-systems game. $N$ groups exist, each with $M$ players. Resources are allocated to each group. Players iterate $T$ game steps and produce offspring who inherit their decision-making functions $f$ with slight mutations. Dashed arrows indicate game steps and solid arrows denote generational turnover. (B) A single game step. The resource grows naturally. Players decide harvesting actions through their $f$ by observing the resource abundance, their own wealth, and the average wealth within the group.}
    \label{fig:DS_game_Mplayer_model_scheme}
\end{figure}

\section*{Model}
We present a common-pool resource management model that formalises the tragedy of the commons: harvesting yields an immediate gain, whereas delaying harvest allows the resource to regenerate with a bonus \cite{hardin1968tragedy}.
This model generalises an earlier two-player model to an $M$-player setting \cite{itao2025self}.

We consider \(N\) groups of \(M\) players that share renewable resources and interact for \(T=1000\) game steps in each generation (Fig. \ref{fig:DS_game_Mplayer_model_scheme}). A player may represent an individual or a collective decision-making unit such as a household or lineage. In each group, the resource abundance \(x(t)\) and player wealth \(\{y_i(t)\}_{i=1}^M\) (i.e., accumulated payoff subject to decay) change across game steps.

At each game step (Fig. \ref{fig:DS_game_Mplayer_model_scheme}(B)), the resource level $x(t)$ changes according to the logistic function $r(x)=x+\alpha (x-x^{2})$, where $\alpha>0$ denotes the intrinsic growth rate. In the following analysis, we set $\alpha=1$, yielding $r(x)=2x-x^{2}$. This form accelerates growth when the resource is scarce and asymptotically approaches $x=1$.

Each player $i$ makes a harvesting decision $h_i(t)$ at each game step $t$ based on the resource level $x(t)$, the player's own wealth $y_i(t)$, and the group-average wealth $\ev{y(t)}$:
\begin{equation*}
h_i(t)=f_i(x(t),y_i(t),\ev{y(t)})=\chi\bigl(x(t)+S_i My_i(t)+A_iM\ev{y(t)}\bigr),
\end{equation*}
where the threshold function $\chi$ is defined as $\chi(z)=1$ if $z>0$ and $\chi(z)=0$ otherwise. Then, \(h_i(t)=1\) denotes harvesting, and \(h_i(t)=0\) denotes waiting. The two strategy parameters \((S_i,A_i)\) quantify sensitivity to self wealth and group-average wealth, respectively.

Because resource abundance is independent of $M$, individual wealth scales as $M^{-1}$. Multiplying both wealth inputs by $M$ keeps the scale of the decision parameters comparable across group sizes. We use a binary harvest-or-wait decision to represent users who commit their harvesting effort in separate periods and can perform other tasks while waiting.

Note that each player's decision rule uses only two parameters rather than $M$ player-specific coefficients. In principle, one could introduce a matrix $W_{ij}$, where $W_{ij}$ captures the effect of player~$j$'s wealth on player~$i$'s decision (see Fig. S3 and \cite{akiyama2002dynamical}). Observing such individual-level information would, however, be prohibitively costly. We therefore restrict social information to the anonymous group-average wealth $\ev{y(t)}$, so that $f_i$ is specified by $(S_i,A_i)$.

\begin{table}[bt]
    \caption{Elements in the model. Elements are categorised as follows: variables $(x,\ y,\ h)$ change at each game step, evolvables $(f,\ S,\ A)$ change over generations, and fixed parameters $(M,\ N,\ \alpha,\ \beta,\ \kappa,\ \mu)$ are held constant within a given simulation. The values of variables and evolvables represent their possible ranges, while the values of fixed parameters indicate those used in the simulations in this paper.}
    \label{table:DS_game_params}
    \centering
    \begin{tabular}{lll}
      & explanation & value \\\hline
      $x$ & The resource abundance & $0\le x \le 1$ \\
      $y$ & The wealth of the player & $y \ge 0$ \\
      $h$ & The action of the player & $0/1$ \\\hdashline
      $f$ & Decision-making function & \\
      $S$ & Weight for the self wealth & $-\infty < S < \infty$ \\
      $A$ & Weight for the average wealth & $-\infty < A < \infty$ \\\hdashline
      $M$ & Number of players in the group & $2 \le M \le 128$ \\
      $N$ & Number of groups in the system & $3 \le N \le 1000$ \\
      $\alpha$ & Resource growth rate & $1$ \\
      $\beta$ & Harvesting fraction & $0.9 / M$ \\
      $\kappa$ & Decay rate of the wealth & $0.25$ \\
      $\mu$ & Mutation rate of $S$ and $A$ & $0.1$
    \end{tabular}
\end{table}

After harvesting, the resource level is updated based on
\begin{equation*}
x(t+1)=r\bigl(x(t)\bigr)\bigl(1-\beta\sum_{i=1}^{M} h_i(t)\bigr).
\end{equation*}
Each harvester gets a fraction $\beta$ of the available resource, so the payoff is $p_i(t) = \beta h_i(t) r(x(t)).$ We set $\beta = 0.9/M$ to normalise the resource use across group sizes.
This normalisation keeps the maximum aggregate fraction harvested per step, $\beta M$, fixed at $0.9$. The results are qualitatively robust across different values of $\beta M$ (Fig.~S1). This can be interpreted as scaling the available resource base with group size, thereby maintaining constant population density and isolating the effect of group size from that of increased aggregate harvesting pressure.
Then, the wealth of player $i$ is updated according to $y_i(t + 1) = (1 - \kappa) y_i(t) + p_i(t),$ where $\kappa$ ($0 < \kappa < 1$) denotes the per-step decay, or consumption, rate. Thus, a player's wealth declines in proportion to its current level, unless new resources are harvested. Here, we set $\kappa = 0.25$.

Individual fitness is defined as the time-averaged wealth after excluding the first 10\% of the game as an initial transient, $\bar y_i = \sum_{t=0.1T}^{T-1} y_i(t)/0.9T$. Excluding the initial transient reduces selection for short-lived gains obtained before resource depletion. We examine other fitness measures in the Supplementary Text.
The total fitness within a group is bounded above by \(\alpha/4\kappa = 1\) (see Supplementary Text for derivation). Accordingly, the group's total fitness provides a size-independent measure of resource-use efficiency.

After each generation, $NM$ offspring are produced. Each player is selected as a parent with probability proportional to its fitness. The resulting offspring counts approximately follow a Poisson distribution with the mean of fitness relative to the population-wide mean.
Offspring inherit strategy parameters \((S_i,A_i)\) with Gaussian mutation of mean $0$ and variance $\mu^2$ added to each parameter. Then, they are randomly reassorted into new groups with fresh resources in the next generation. Here, we set $\mu = 0.1$.
Each simulation was run for 3000 generations.

One can always outperform other group members by harvesting at every step. However, in a population subdivided into $N$ groups, selection favours players who attain higher fitness than members of other groups, rather than merely outperforming their intragroup rivals. This introduces multilevel selection, whose balance is set by $N$ and $M$. We examine how the conditions for sustainability depend on these two parameters.

The full list of model components is provided in Table~\ref{table:DS_game_params}. The decision parameters are initialised as $S=A=0$. At the start of each generation, the state variables are reset to $x(0)=0.1$ and $y_i(0)=(1+\eta_i)/M$, where $\eta_i$ denotes Gaussian noise with mean $0$ and variance $0.01$.
We assess sensitivity to parameter values and alternative model specifications in the Supplementary Text. While several extensions produce consistent results, others identify different conditions for the evolution of sustainable institutions (see Figs.~S1--S3 and S15--S16).

\section*{Results}
We first present evolutionary simulations showing that sustainable resource use and turn-taking arise endogenously. Then, we quantify how moderate strategic diversity (i.e., the variance in decision parameters within a group), as well as noise, improves resource-use efficiency, equity, and robustness. Finally, we examine how the minimum number of groups required for sustainability, $N_c$, scales with group size $M$.

\subsection*{Self-organisation of institutions}
\begin{figure*}[tb]
  \centering
   \includegraphics[width=\linewidth]{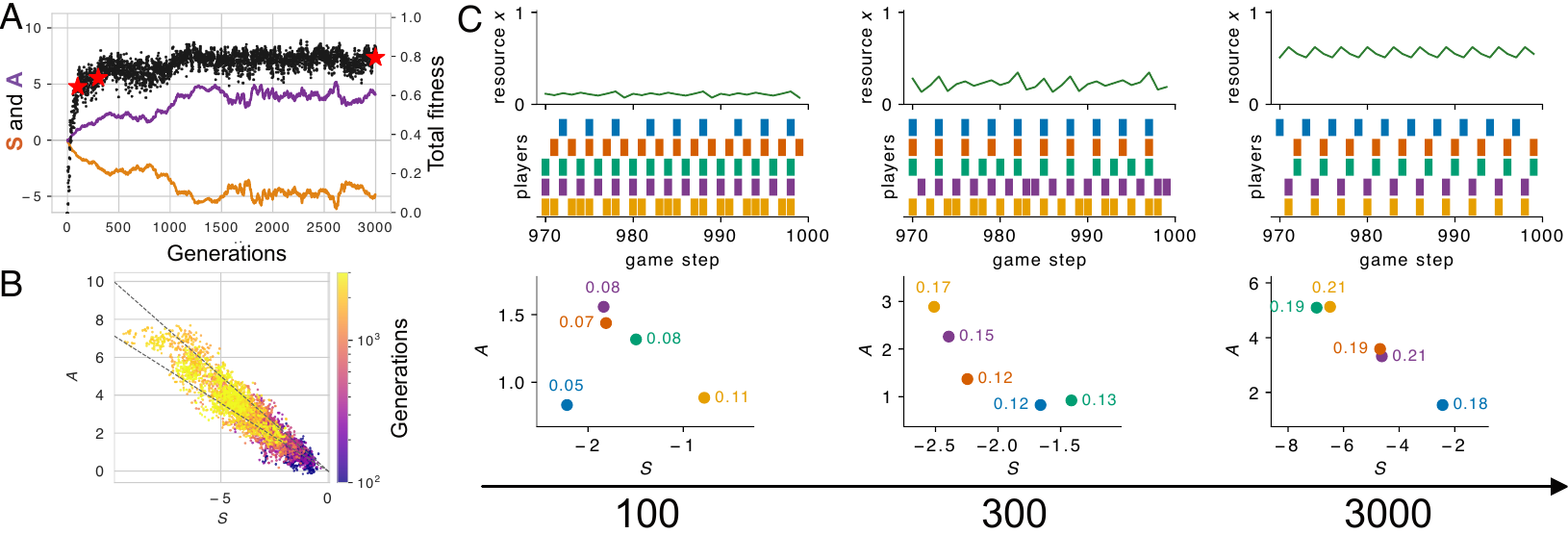}
   \caption{Example of the evolutionary dynamical-systems game with $M = 5$ players and $N = 100$ groups.
(A) Evolution of the population averages of the decision parameters $S$ (orange) and $A$ (purple). Black dots denote the mean total fitness of $N$ groups (right axis). Red stars mark the generations highlighted in (C).
(B) Temporal change in the distribution of decision parameters. Dashed lines show $S/A = -1.1$ and $-1.3$.
(C) Game dynamics during the final 30 steps (steps $970$–$999$). Green curves show resource abundance (top), and coloured squares mark each player's harvesting actions (middle). The bottom panels show each player's $(S,A)$ values and fitness in the corresponding colour. Numbers beneath each panel give the generation index.}
    \label{fig:DS_game_Mplayer_temporal}
\end{figure*}

Through evolution, periodic harvesting patterns emerged (Fig. \ref{fig:DS_game_Mplayer_temporal}). We focus first on \(M=5\) and \(N=100\), but qualitatively similar results are obtained for \(M=3\) and \(10\) (Figs. S4 and S5). Across 100 trials per condition, turn-taking systems supporting sustainability emerged in every trial, although their cycle lengths and the generation at which they appeared varied.

Within a generation, the joint dynamics of resource abundance \(x\) and player wealth \(y_i\) converge to periodic attractors (Fig. \ref{fig:DS_game_Mplayer_temporal}(C)). Because the decision-making functions \(f_i\) are fixed within a generation, identical actions \(\{h_i\}\) are triggered whenever the same combination of \(x\) and \(\{y_i\}\) recurs. The attractors are either limit cycles or a fixed point with \(x(t)\to 0\) under severe overharvesting. We did not observe chaotic dynamics (see Supplementary Text for Lyapunov stability analysis).

Across generations, fitness rises as harvesting becomes less frequent and the resource recovers. In the example of Fig. \ref{fig:DS_game_Mplayer_temporal}(C), harvesting rates in generation 100 are high and vary among players, so frequent harvesters exploit waiting players by overusing the scarce resource. By generation 300, most players harvest less often and the resource recovers, but the remaining variation in harvesting rates still permits exploitation. By generation 3000, harvesting rates have converged within the group, and each player harvests once every three steps. The players form three clusters: (1) blue, (2) orange and green, and (3) purple and yellow. The action order is \(1\rightarrow 3 \rightarrow 2\rightarrow 1 \rightarrow \cdots\). The resource recovers after the single player in cluster 1 harvests and is then depleted by the two-player clusters 3 and 2. Cluster 3, which harvests immediately after cluster 1, therefore obtains the highest fitness.

Thus, within this generation, the four players forming the majority obtain higher fitness than the single player forming the minority. This difference arises from unequal cluster sizes and harvesting order, which result from random initial conditions and symmetry breaking as discussed below, and from differences in $S$ and $A$. Nevertheless, fitness is more evenly distributed than in earlier generations.

This evolutionary trend is driven by shifts in the decision parameters of \(f\) (Fig. \ref{fig:DS_game_Mplayer_temporal}(A, B)). Recall that player \(i\) harvests if \(x + S_i My_i + A_i M\ev{y} > 0\). Across generations, \(S\) decreases, \(A\) increases, and the diversity of \((S,A)\) within the population expands. Throughout this process, successful strategies remain within \(-1.3 < S/A < -1.1\). Players therefore refrain from harvesting when they are wealthy but become more likely to harvest when the group average is high. Negative \(S\) allows periods of waiting during which the resource recovers. Positive \(A\) introduces a collective sanctioning response to make the system robust against selfish overharvesters. A player that harvests too frequently becomes wealthier than the others and raises \(\ev{y}\), which induces the remaining players to harvest more often and thereby depresses resource abundance. Such a response is costly at the group level. Groups in which persistent defection triggers it contribute little to subsequent generations, whereas cooperative groups achieve higher fitness, allowing cooperation to persist. When \(A \simeq -S\), the same rule also generates labour division, as players who have just harvested tend to wait while others act.

Thus, monitoring only anonymised average wealth allows players to respond to the aggregate signature of overharvesting through a collective sanction, without identifying the responsible individual. Both the criterion of cooperation and the sanctioning response to its violation, two defining components of an institution, emerge endogenously when $S < 0$, $A > 0$, and $-1.3 < S/A < -1.1$. The emergent institution follows a simple principle: \emph{wait when wealthy, harvest when below average}. It supports both sustainable resource use and an equitable division of labour.
Strategies evolving outside this region lead to overharvesting (Fig.~S6). Moreover, even when players can monitor each group member's wealth separately, evolution converges on the same simple principle (Fig.~S3).

\begin{figure}[tb]
  \centering
   \includegraphics[width=.6\linewidth]{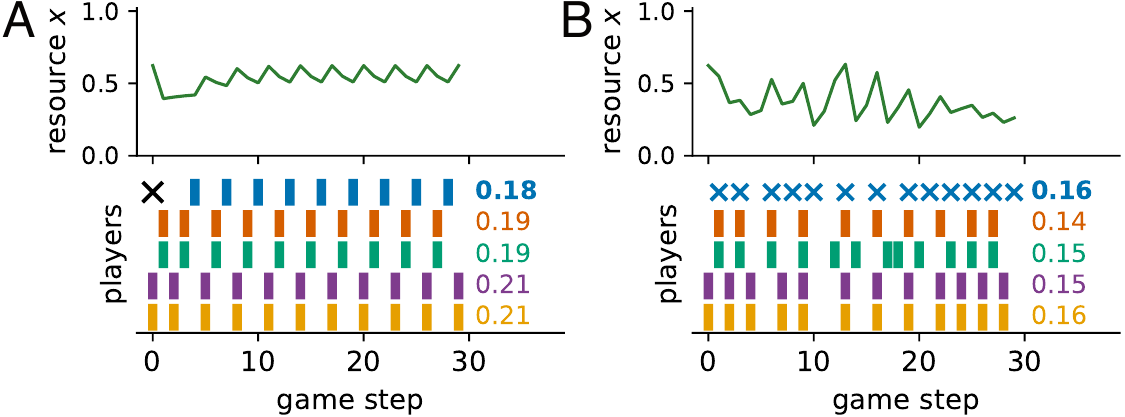}
    \caption{Sanctioning under emergent institutions. We sampled a group from generation 3000 in Fig.~\ref{fig:DS_game_Mplayer_temporal}(C) and ran two follow-up simulations. The panels show the first 30 steps of each continuation from the final state of that group. Green curves show the resource dynamics. Coloured squares and values indicate the harvesting actions and fitness of players, respectively.
    (A) Temporal sanctioning after player~1 makes an unintentional harvest at step 0 (indicated by $\times$), without modifying decision parameters.
    (B) Severe sanctioning against a selfish mutant created by lowering player~1's harvesting threshold to \(x + S_1 My_1 + A_1 M\ev{y} + 2 > 0\) throughout the 1000-step continuation.}
    \label{fig:DS_game_Mplayer_punish}
\end{figure}

\begin{figure*}[tb]
  \centering
   \includegraphics[width=.75\linewidth]{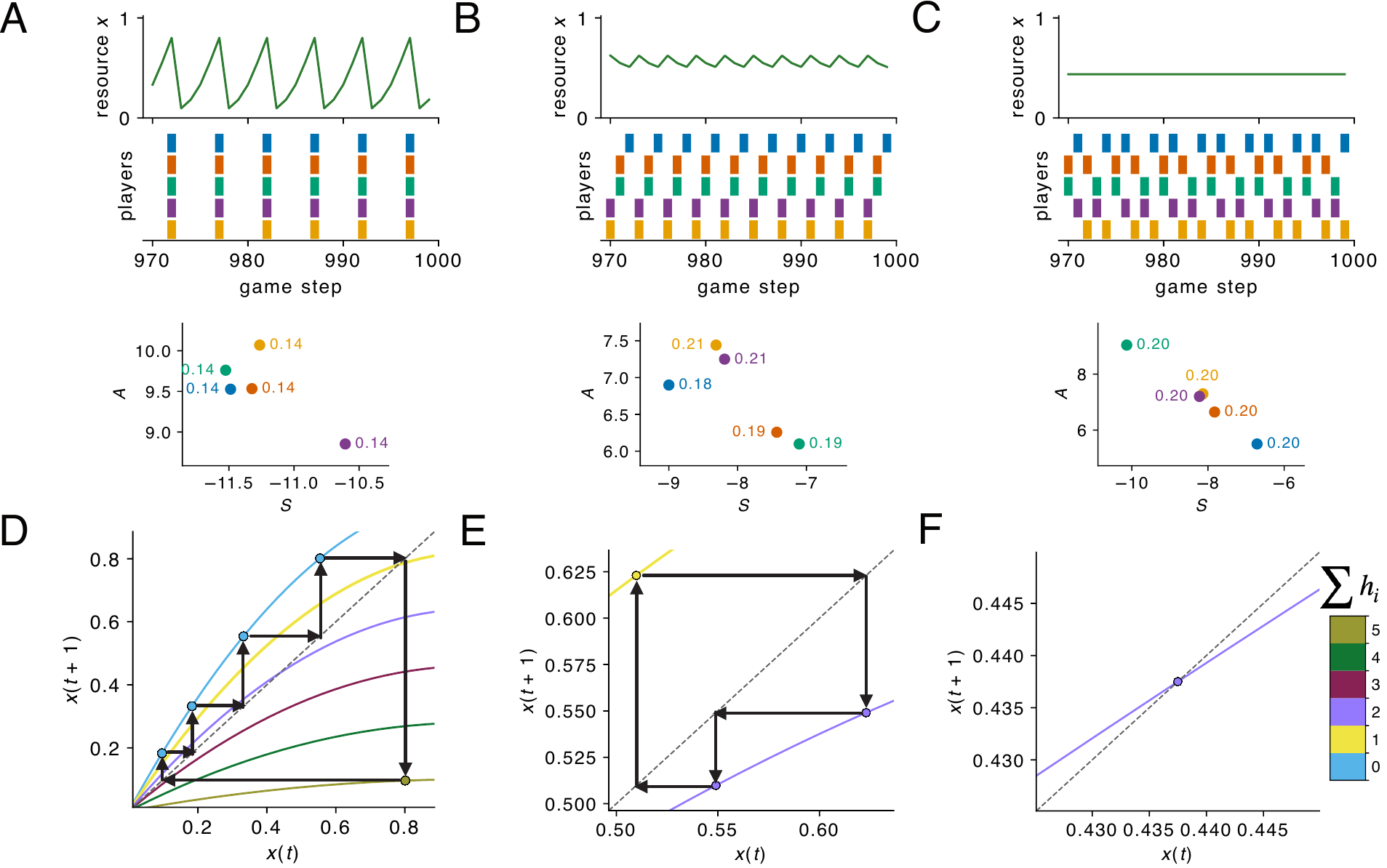}
   \caption{Modes of the game dynamics.
(A–C) Dynamics over the final 30 game steps. Green curves show the resource dynamics, coloured squares indicate the harvesting actions of players, and coloured circles show players' $(S, A)$ values with their fitness.
(D–F) Return maps of the resource dynamics. The next abundance $x(t+1)$ depends on the current abundance $x(t)$ and the number of harvesters $\sum h_i(t)$. Coloured lines indicate the possible updates $x(t+1) = r(x(t))(1-\beta\sum h_i(t))$ for different $\sum h_i(t)$, as shown by the colour bar at the lower right.}
    \label{fig:DS_game_Mplayer_modes}
\end{figure*}

Figure~\ref{fig:DS_game_Mplayer_punish} illustrates the sanctioning mechanism encoded by the evolved institution. We selected one group from generation 3000 in Fig.~\ref{fig:DS_game_Mplayer_temporal} and ran two follow-up simulations. In Fig.~\ref{fig:DS_game_Mplayer_punish}(A), player~1 makes a one-off erroneous harvest at \(t=0\) without changing its decision parameters. The other players transiently increase their harvesting, while player~1 decreases its harvesting, and the original limit cycle is restored within 5 steps. In Fig.~\ref{fig:DS_game_Mplayer_punish}(B), player~1 is converted into a selfish mutant by relaxing its threshold to \(x + S_1 My_1 + A_1 M\ev{y} + 2 > 0\). Adding the constant term \(2\) to the left-hand side causes player~1 to harvest in more contexts, leading to persistent cheating. In response, the other players harvest more frequently.
Relative to the original dynamics in Fig.~\ref{fig:DS_game_Mplayer_temporal}(C), this response lowers the fitness of the mutant and the other players alike, penalising the mutant at a cost to the residents. Given persistent cheating, sanctioning through overharvesting is a damage-limiting response, because continued restraint would allow the mutant to appropriate most of the resource, leaving little for the other players. The cost is mediated by depletion of the shared resource and borne by the group as a whole rather than by the sanctioning players alone. This precludes free-riding on others' enforcement and thereby distinguishes the mechanism from conventional costly punishment.
Although the game dynamics do not become periodic within 30 steps, they approach a different limit cycle with a lower mean resource level. Thus, the evolved institution tolerates occasional mistakes but imposes severe sanctions on persistent defection.

\subsection*{Spontaneous turn-taking in resource use}
Game dynamics converge to distinct ``modes'', defined as periodic patterns in resource abundance and players' wealth \cite{itao2025self}. Fig. \ref{fig:DS_game_Mplayer_modes} displays three representative cases: (A) synchronised harvesting, (B) labour division among three clusters, and (C) complex turn-taking. In each mode, players spontaneously form clusters of coordinated actions, a phenomenon unique to collective interactions. In Fig. \ref{fig:DS_game_Mplayer_modes}(A, B), the wealth dynamics are identical for all players within the same cluster. In Fig. \ref{fig:DS_game_Mplayer_modes}(C), the players' action sequences and wealth dynamics converge to a common attractor, but with phase shifts among players.
These modes illustrate that, even under the same guiding principle of \emph{wait when wealthy, harvest when below average,} subtle shifts in the comparison thresholds generate a large repertoire of complex turn-taking patterns.

\begin{figure*}[tb]
  \centering
   \includegraphics[width=\linewidth]{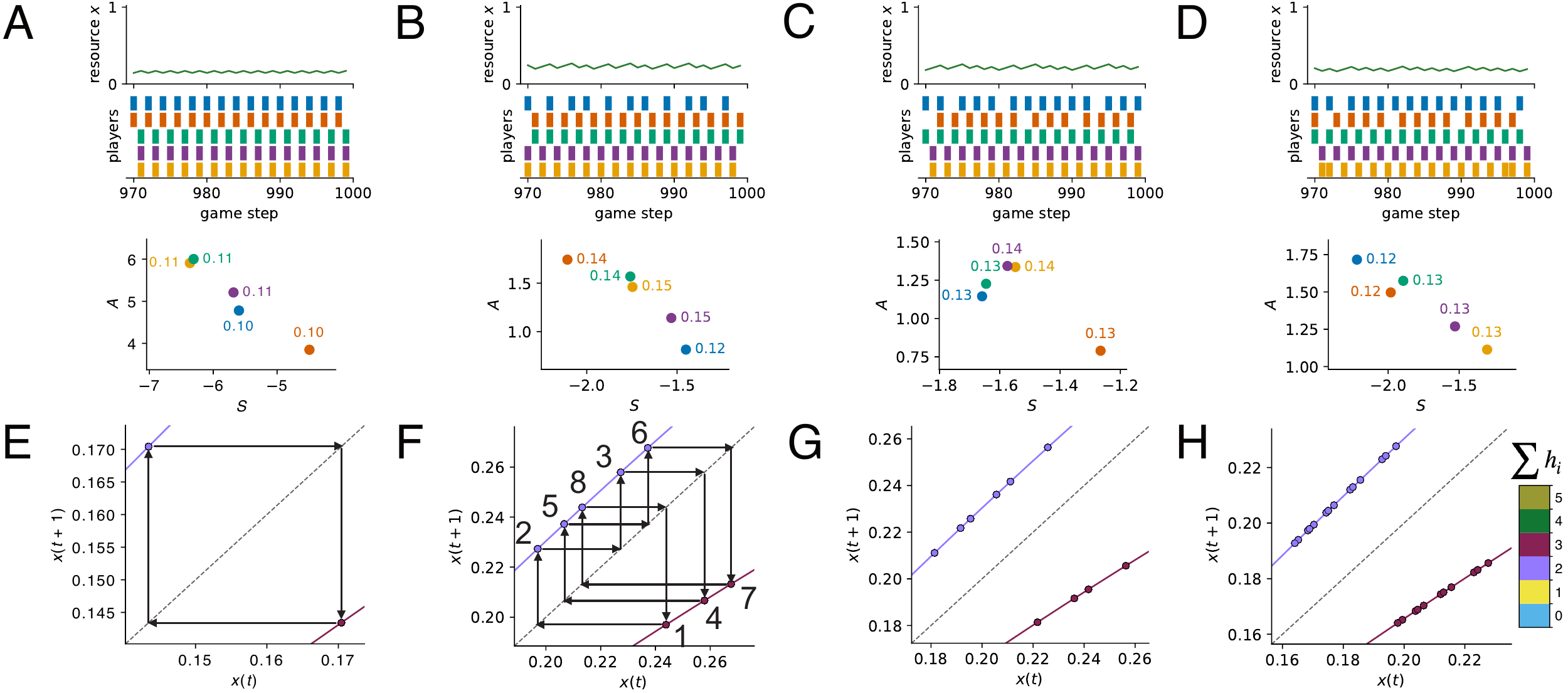}
   \caption{Period elongation of modes.
(A–D) Dynamics over the final 30 game steps. Green curves show the resource dynamics, coloured squares indicate the harvesting actions of players, and coloured circles show players' $(S, A)$ values with their fitness values.
(E–H) Return maps of the resource dynamics. Coloured lines indicate $r(x(t))(1-\beta\sum h_i(t))$ for different $\sum h_i(t)$, as shown by the colour bar at the lower right. In (E) and (F),
arrows trace the resource trajectory, but they are omitted in (G) and (H) for clarity.}
    \label{fig:DS_game_Mplayer_modes23}
\end{figure*}

The sequence of players' actions uniquely determines the dynamics of both the resource and the players' wealth. The resource follows the return maps in Fig. \ref{fig:DS_game_Mplayer_modes}(D--F): the next resource level \(x(t+1)\) depends on the current resource level \(x(t)\) and the number of harvesters \(\sum_i h_i(t)\) according to \(x(t+1)=r\bigl(x(t)\bigr)\bigl(1-\beta\sum_i h_i(t)\bigr)\). Given an action sequence, the resource dynamics follow a consistent orbit on the corresponding return map.
The coupled dynamics of actions and resources then determine the players' wealth dynamics.

Cluster formation and labour division depend on both the decision parameters and stochastic initial conditions. At the start of each generation, player wealth is initialised as $y_i(0) = (1+\eta_i) / M$, where $\eta_i$ is a small Gaussian noise.
Wealth dynamics then follow $y_i(t+1)=(1-\kappa)\,y_i(t) + \beta\,h_i(t)\,r\bigl(x(t)\bigr).$
Players with identical action sequences receive identical payoffs, so differences in their wealth decay by a factor of $1-\kappa$ at each step, driving their trajectories toward identical attractors.
As a harvest is triggered when $x(t)+S_i My_i(t)+A_i M\ev{y(t)} > 0,$ if $|S_i|$ and $|A_i|$ are small, decisions depend largely on the resource level $x(t)$, which promotes synchronised harvesting.

In contrast, when $|S|, |A| \gg 1$ and $S_i \simeq -A_i$ (as observed in later generations in Fig. \ref{fig:DS_game_Mplayer_temporal}), the decision threshold reduces to $S_i(y_i - \ev{y}) > 0$, so actions are governed mainly by relative wealth. Then, synchronised harvesting is unstable. Poorer players harvest while wealthier ones wait, producing turn-taking and thus labour division. Wealth differences are amplified across turn-taking clusters, whereas they decay within clusters, yielding distinct behavioural clusters. The stability of clusters depends on the magnitude of $|S|$ and $|A|$.
For two equally sized clusters \(A\) and \(B\), with \(Y_A=\frac{2}{M}(y_1+\cdots+y_{M/2})\) and \(Y_B=\frac{2}{M}(y_{M/2+1}+\cdots+y_M)\), and with \(y_1 \simeq Y_A\), one has \(x + S_1 My_1 + A_1 M\ev{y} \simeq x + \frac{S_1}{2}My_1 + \frac{A_1}{2}MY_B\). Thus, within-cluster synchrony is stable when \(|S|/2\) and \(|A|/2\) are sufficiently large. More generally, finer clustering requires larger parameter magnitudes (cf. spontaneous clustering in globally coupled chaotic systems \cite{kaneko1990clustering}).
Cluster membership depends on both the parameter values and the initial conditions. When the parameters are nearly identical, membership is determined largely by chance.

\subsection*{Benefits of strategic diversity and stochasticity}
Emergent modes display diverse clustering patterns. Fig. \ref{fig:DS_game_Mplayer_modes23} shows modes in which either two or three players harvest at each step. The elementary mode has period 2 with two alternating clusters (Fig. \ref{fig:DS_game_Mplayer_modes23}(A)). Longer periods arise when some players join or leave these clusters intermittently, producing period-8, period-20, and period-50 cycles (Fig. \ref{fig:DS_game_Mplayer_modes23}(B--D)). Thus, subtle shifts in \((S_i,A_i)\) can produce complex turn-taking patterns with cluster membership changing over time.

Longer-period modes can improve both efficiency and equity. In the elementary mode, the group-average harvesting rate is \(0.500\) and the total fitness is \(0.536\), while the longer-period modes in Fig. \ref{fig:DS_game_Mplayer_modes23}(B--D) achieve \(0.475\) (\(0.709\)), \(0.480\) (\(0.677\)), and \(0.488\) (\(0.622\)), respectively. Because the analytically optimal average harvesting rate is \(0.37\) (see Supplementary Text), longer periods allow finer adjustment of harvesting rate. They also permit cluster membership to change over time, i.e., ``turn-taking of turns'', so that majority and minority roles alternate and fitness becomes more equal within a group. Hence, longer periods are useful not in themselves, but because they enable finer harvest control and higher-order alternation between cluster roles.

\begin{figure*}[tb]
  \centering
   \includegraphics[width=\linewidth]{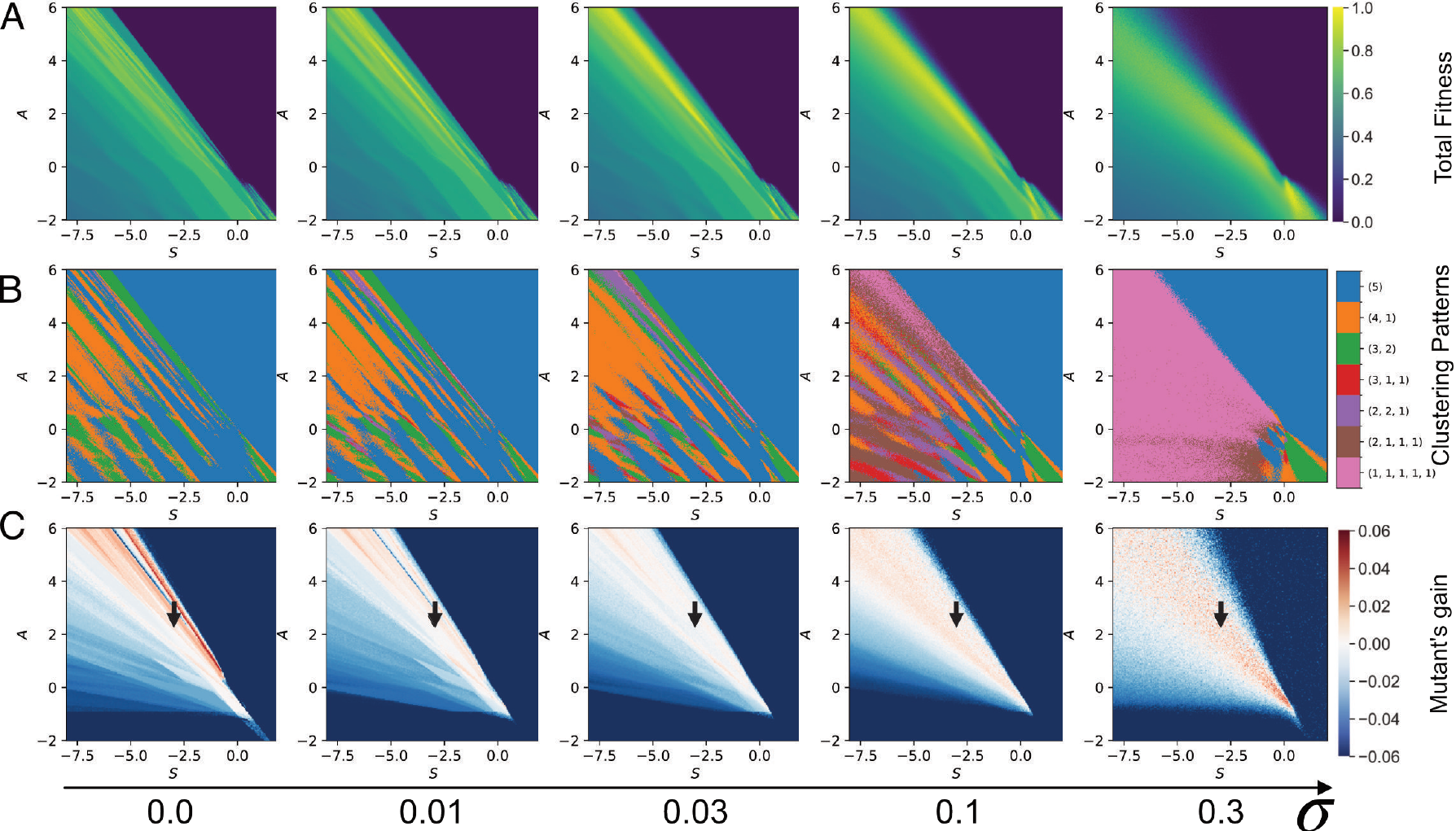}
   \caption{Decision‐parameter dependence of fitness, clustering, and robustness under varying strategic diversity.
(A) Total fitness in the group.
(B) Clustering patterns of players' action sequences; for example, ``(3, 2)'' denotes two clusters with three and two players. In panels (A) and (B), five strategies are generated by multiplying Gaussian perturbations with mean $1$ and variance \(\sigma^{2}\) to each \((S, A)\).
(C) Evolutionary robustness. Four resident strategies (distributed around the arrowhead at $(S, A)=(-3, 2.5)$) are fixed, while the mutant's strategy sweeps the parameter space. Colours show the difference between mutant fitness and mean fitness in an all-resident group at the same $\sigma$. Red dots indicate parameter regions that can invade residents. Each dot summarises $100$ trials for the corresponding \((S,A)\).}
    \label{fig:DS_game_Mplayer_phase}
\end{figure*}

The periods of the modes lengthen as the decision parameters $(S_i, A_i)$ become more heterogeneous, with such heterogeneity expanding through evolution (Fig. \ref{fig:DS_game_Mplayer_temporal}(B)). We refer to the within-group variance in decision parameters as ``strategic diversity''. When $S_i$ and $A_i$ differ within a cluster, players facing the same triplet $(x, y_i, \ev{y})$ may select different actions. This heterogeneity destabilises synchronisation, promotes turn-taking, and thereby generates longer-period dynamics. To clarify how game dynamics depend on the strategic diversity, we ran simulations in which players' parameters are randomly distributed around prescribed values (Fig. \ref{fig:DS_game_Mplayer_phase}). For each pair $(S, A)$, the parameters of the $M = 5$ players were drawn as
$(S_i, A_i) = \bigl(S\zeta_i^{(1)},\, A\zeta_i^{(2)}\bigr),$
where $\zeta_i^{(1)}$ and $\zeta_i^{(2)}$ are independent Gaussian noise with mean $1$ and variance $\sigma^{2}$. The standard deviation $\sigma$ quantifies the strategic diversity. In the leftmost panels of Fig. \ref{fig:DS_game_Mplayer_phase} ($\sigma = 0$), all players adopt identical strategies; positive $\sigma$ values introduce heterogeneity into the population. (In the two-player model, mode dependence was illustrated by ``iso-mode regions'' \cite{itao2025self}. In the collective setting, however, the number of modes increases due to longer-period dynamics. Thus, we summarise outcomes by total fitness and clustering patterns instead.)

First, without strategic diversity, Fig. \ref{fig:DS_game_Mplayer_phase}(A) illustrates the
dependence on the decision parameters, where total fitness peaks along the narrow band where $-1.3 < S/A < -1.1$. Fitness is minimal in the upper-right region, where large values of $S$ and $A$ drive excessive harvesting and deplete the resource, and it also declines in the lower-left region, where small $S$ and $A$ suppress harvesting and leave the resource underutilised. Hence, total fitness is maximised when $S$ and $A$ are balanced, which prevents both over- and underharvesting.

A moderate level of strategic diversity raises group fitness by enabling the formation of additional clusters in the action sequence. Fig. \ref{fig:DS_game_Mplayer_phase}(B) reveals that spontaneous labour division arises even when all players share identical decision parameters; however, without diversity ($\sigma = 0$) the population supports at most two clusters, and the dynamics settle into short cycles such as the elementary mode in Fig. \ref{fig:DS_game_Mplayer_modes23}(A). Introducing a moderate $\sigma$ permits longer-period dynamics with more clusters, allowing finer regulation of the average harvesting rate and thereby improving fitness. By contrast, large $\sigma$ values generate excessive variation in harvesting rates and reduce the efficiency of resource use.
Consistent with this pattern, Fig. S7 shows that the period lengthens as $\sigma$ increases, while fitness attains its maximum at an intermediate $\sigma$ when $M \ge 3$. In the special case $M = 2$, the system supports only synchronisation or a simple labour division, so more elaborate ``turn-taking of turns'', as in Fig. \ref{fig:DS_game_Mplayer_modes23}(B–D), cannot arise, and the benefits of strategic diversity remain minimal.

Strategic diversity also enhances evolutionary robustness against mutant strategies. In Fig. \ref{fig:DS_game_Mplayer_phase}(C), four resident strategies are fixed around \((S,A)=(-3, 2.5)\), each perturbed by variance \(\sigma^2\), while a fifth strategy sweeps the parameter plane. We measure the mutant's gain as the difference between its fitness when competing with the residents and the mean fitness obtained when all five strategies are residents. A negative gain means that the mutant does best by matching the residents rather than deviating from them. We therefore use the extent of the parameter region around the residents in which mutant gain remains negative as a measure of the residents' evolutionary robustness (cf. a more precise definition for the two-player case is given in \cite{itao2025self}).

Evolutionary robustness is maximised at intermediate \(\sigma\). When \(\sigma=0\), identical residents can be exploited if the mutant happens to escape sanctioning. Because moderate diversity makes resident responses heterogeneous and sanctioning by even a subset of residents can erase the mutant's advantage, the mutant gains little at intermediate \(\sigma\). Fig. S8 shows the corresponding mechanism: low \(\sigma\) often leads to a split into two clusters (i.e., a mutant versus residents), whereas higher \(\sigma\) induces sanctioning by some residents. Diversity improves robustness because a mutant must evade several distinct responses simultaneously rather than exploit a single resident strategy.

Thus, Fig. \ref{fig:DS_game_Mplayer_phase} highlights two advantages of strategic diversity:
(i) it raises resource-use efficiency by permitting more complex turn-taking, and
(ii) it enhances evolutionary robustness by diversifying sanctioning decisions.
Both effects are consistently observed for $M \ge 3$ (see Figs. S10 and S11).

Finally, we introduce stochasticity in resource dynamics and peer monitoring as
\begin{align}
r(x) &= (2x-x^2)(1+\eta),\\
h_i(t) &= \chi\!\bigl(x(t)+S_i My_i(t)+A_iM\ev{y(t)}(1+\eta_i)\bigr),
\end{align}
where \(\eta\sim\mathcal{N}(0,\sigma_r^2)\) and \(\eta_i\sim\mathcal{N}(0,\sigma_p^2)\) denote environmental and monitoring noise, respectively. The latter can also be interpreted as stochasticity in decision-making, because it is equivalent to introducing stochastic fluctuations in the effective decision threshold. Fig. S12 shows the dependence of the total fitness on noise levels. Total fitness peaks at intermediate noise levels, around \(\sigma_r=0.02\) and \(\sigma_p=0.1\). Moderate noise induces irregular harvesting that mimics the beneficial effects of strategic diversity and facilitates higher-order alternation among clusters.

\subsection*{Conditions for the emergence of institutions}
\begin{figure}[tb]
  \centering
   \includegraphics[width=\linewidth]{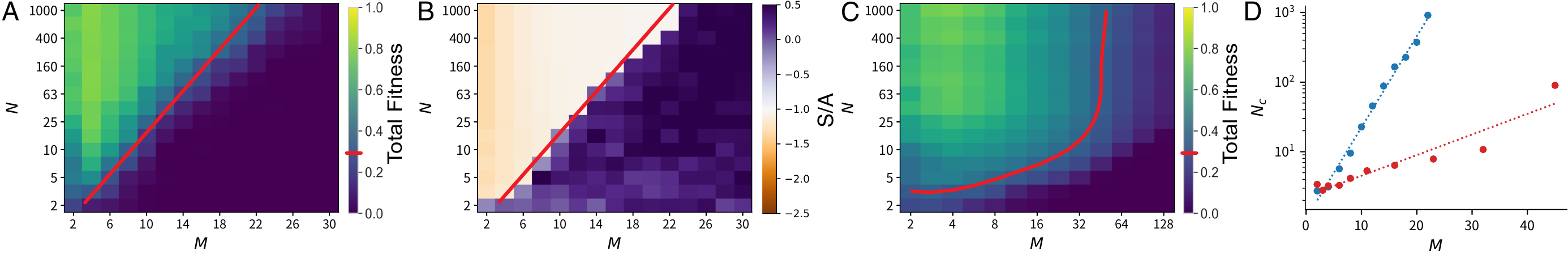}
\caption{Conditions for sustainability. (A, C) Mean total fitness as a function of group size $M$ and number of groups $N$ when (A) $M$ remains fixed or (C) is multiplied by $\sqrt{2}$ every $1000$ generations. (B) The median evolved ratio $S/A$ under fixed $M$. Red lines and curve in (A--C) mark the boundary between sustainability, defined as total fitness $>0.3$, and overharvesting. (D) The minimum number of groups required for sustainability, $N_c$, under fixed group size (blue) and gradual expansion (red). Points show the estimated boundaries, and dashed lines show exponential fits, $N_c\propto\exp(0.3M)$ (blue) and $N_c\propto\exp(0.07M)$ (red).}
    \label{fig:DS_game_Mplayer_phase_NM}
\end{figure}

The risk of the tragedy of the commons increases with group size $M$ and decreases with the number of groups $N$. Each additional member can introduce a strategy incompatible with sustainable harvesting. Because sanctioning through overharvesting lowers the fitness of members of a group containing such a cheater, players in cooperative groups leave more offspring. A larger $N$ provides more opportunities for cooperative groups to form, whereas a larger $M$ makes them rarer. The minimum number of groups required for sustainability, $N_c$, therefore increases with $M$.

Figure~\ref{fig:DS_game_Mplayer_phase_NM}(A) shows mean total fitness over $100$ trials for each combination of $M$ and $N$. Although the upper bound of total fitness is $1$ for every $M$, sustainable populations become increasingly rare as $M$ grows. Fig.~\ref{fig:DS_game_Mplayer_phase_NM}(B) shows that the fitness boundary at $0.3$ coincides with an abrupt change in the evolved decision rule. In sustainable populations, $S/A$ remains near $-1$, consistent with the principle \emph{wait when wealthy, harvest when below average}, whereas this relationship is lost below the boundary.

The persistence of the principle \emph{wait when wealthy, harvest when below average} across group sizes (Figs.~S9--S11) suggests that populations need not rediscover the rule whenever $M$ changes. We therefore examined gradual demographic expansion. In the simulation shown in Fig.~S13, $M$ is multiplied by $\sqrt{2}$ every $1000$ generations by increasing the total number of offspring while holding $N$ constant. The decision parameters $(S,A)$ adjust as the population expands, allowing sustainability to persist at group sizes for which populations evolving directly at fixed $M$ usually collapse.

Figure~\ref{fig:DS_game_Mplayer_phase_NM}(A), (C), and (D) show that $N_c$ increases approximately exponentially with $M$ in both scenarios. We estimate $N_c$ by interpolating threshold crossings in $\log N$ and fit $\log N_c=aM+b$ by least squares. The fitted rate drops from $a\simeq0.3$ under fixed group size to $a\simeq0.07$ under gradual expansion. A simple approximation explains this exponential scaling. If each individual achieves a sustainable strategy with probability $p$, a group contains a sustainable combination of $M$ strategies with probability $p^M$. Requiring at least one such group gives $Np^M>1$, so that $N_c\simeq p^{-M}=\exp(-M\log p)$. The fitted rates correspond to effective values of $p\simeq0.74$ under fixed group size and $p\simeq0.93$ under gradual expansion. This difference is consistent with the distinction between establishing sustainable strategies de novo and retaining them as groups expand. Under gradual expansion, however, total fitness levels off for $N \geq 40$ at a value that declines with $M$, so that no $N$ reaches the threshold for $M \geq 64$. The exponential scaling over the fitted range persists across the other tested total-fitness thresholds (Fig.~S14).

The robustness analyses identify both general features and limiting assumptions (Figs. S15 and S16). Sustainable resource use remains possible when fitness is defined by total harvest, decision rules are represented by RNNs, players can observe either the wealth of every group member or the group maximum only, alternative schedules of group-size expansion are used, the threshold decision is replaced by a stochastic sigmoid, or mutations are heavy-tailed. Persistent group structure allows sustainability over a wider range of group sizes. By contrast, including the full initial transient, removing peer information, or keeping the individual harvesting fraction fixed generally restricts the range supporting sustainability. A population that first evolved under transient exclusion can nevertheless maintain sustainability when fitness is subsequently averaged over all steps. No sustainable outcome was observed when resources were carried over instead of being refreshed in each generation.

\section*{Discussion}
\subsection*{Self-organised turn-taking}
By extending evolutionary dynamical-systems games to $M$-player common-pool resource management, we showed that sustainable governance can emerge even when players observe only anonymised group-level information. The evolved decision rule can be summarised as \emph{wait when wealthy, harvest when below average}. This principle induces players to self-organise into temporal clusters, take turns harvesting, and sanction persistent overharvesting, thereby supporting sustainability, equity, and robustness.
The first component lets the resource regenerate, whereas the second serves as a built-in sanction that penalises overharvesters. Persistent overharvesting raises group-average wealth and induces others' harvesting that depresses the resource level, thereby reducing the gains from defection.
In this sense, sanctions are not imposed externally and need not target identified violators. Emergent sanctioning was progressive: occasional mistakes were tolerated after transient sanctions, whereas persistent defection elicited increasingly severe sanctions.
Thus, both the criterion for cooperation and the sanctioning response to its violation emerge endogenously. Together, they constitute a self-organised institution.

The simple principle generates a large repertoire of resource use patterns. When decisions depend mainly on resource abundance, players synchronise their actions, corresponding to seasonal closure for resource use. When decisions depend mainly on relative wealth, poorer players harvest while wealthier players wait, producing a labour division among clusters of players. Small changes in the thresholds set by $S$ and $A$, together with within-group variation in these parameters, then generate longer-period rotations in which individuals switch among clusters. Thus, turn-taking is not an externally prescribed sequence. It emerges through the interaction between resource regeneration and evolving decision rules.

Moderate strategic diversity and stochasticity make this coordination more effective. Homogeneous groups can already form turn-taking clusters, but unequal cluster sizes leave persistent fitness differences between majority and minority roles. Heterogeneous strategies allow cluster membership to change over time, creating a higher-order \emph{turn-taking of turns} that improves resource-use efficiency and equity. Heterogeneity also makes it harder for mutants to exploit resident populations by diversifying sanctioning decisions. Moderate noise in resource dynamics or peer monitoring plays a similar role by inducing irregular harvesting, whereas excessive diversity or noise disrupts efficient use. These results suggest that heterogeneity and stochasticity can be beneficial when they provide flexibility without destroying the shared principle of action, consistent with empirical work showing that cognitive diversity can improve commons governance \cite{baggio2019importance}.

The conditions for sustainability depend on demographic history. The minimum number of groups increases approximately as $N_c\propto\exp(0.3M)$ under fixed group size and as $N_c\propto\exp(0.07M)$ under gradual expansion. Gradual expansion therefore makes sustainability more feasible. These different rates reflect the distinction between establishing sustainable strategies and maintaining them as groups grow. Because the principle \emph{wait when wealthy, harvest when below average} persists across group sizes, expanding populations can adjust an existing rule rather than rediscover sustainable strategies.

\subsection*{Relevance to institutional theory}
These findings complement institutional analyses of common-pool resource governance. Empirical work has long emphasised that locally devised rules, user autonomy, and monitoring are central to successful resource management \cite{ostrom1990governing, ostrom2009general, cox2010review}. Our model provides a mechanistic account of how such rules can arise before they are formally codified. No action sequence is \emph{a priori} cooperative. The criteria of cooperation emerge from the coupled dynamics of resource growth, harvesting, demographic history, and evolving decision-making functions. This suggests that institutions may originate from simple context-dependent principles before being expressed as explicit rules.

Many community-based institutions use rotational access to govern common-pool resources \cite{ostrom1990governing, lobe2004padu}. The present results suggest that such rotations can arise from the general principle of \emph{wait when wealthy, harvest when below average}. This distinction is important because a fixed rotation may become fragile when group size or resource conditions change. A decision principle that dynamically regenerates rotations can provide flexibility in action patterns and robustness against changes in group size or resource conditions. This view is also consistent with field and laboratory studies showing that turn-taking can self-organise without detailed formal prescriptions \cite{cason2013learning}.

The sanctioning mechanism broadens the interpretation of institutional enforcement. Existing analyses often focus on targeted sanctions against identifiable violators \cite{ostrom1990governing, cox2010review}. By contrast, the emergent sanctions in our model are triggered by elevated group-average wealth, which signals persistent overharvesting without identifying specific violators. Other players then harvest more, degrading the resource and reducing the payoff of persistent overharvesters. Although this response is costly for the affected group, it leaves persistent overharvesters with lower fitness than they would obtain by cooperating. Empirical cases suggest that users sometimes respond to perceived overuse by increasing their own extraction \cite{mbaiwa2014community, pons2022trade, farmer2020forecasting, cudney2009lack}. Such responses are often interpreted as failures of governance, but our results suggest that under limited information, they can prevent overharvesters from gaining a fitness advantage. This mechanism is especially plausible when resource use is divided into discrete seasons or patches, so that overharvesting prematurely terminates only a local opportunity and sanctioning players can benefit elsewhere or in later cycles. In this sense, aggregate information can support a rudimentary form of sanctioning, which may later be replaced or reinforced by targeted punishment once monitoring systems become sufficiently elaborated.

Several limitations remain.

First, even under gradual expansion, the number of groups required for sustainability scales approximately exponentially with group size. Furthermore, sustainable resource use was not observed under full resource carry-over. The results therefore apply most directly to settings with renewed resources. Otherwise, collective sanctions through resource depletion may be too costly to sustain cooperation. Targeted sanctions or learning within generations may relax these conditions.
Second, institutional emergence at large $M$ depends on excluding the initial transient from fitness. This is reasonable when the focus is on long-run patterns of resource use, but it limits applicability to short-term settings, in which transient harvesting cannot be ignored.
Third, monitoring itself can be costly. Investments that make individual behaviour observable may coevolve with incentives to conceal cheating, and this information game is not included here.

Overall, the model shows how coarse-grained information can support the self-organisation of institutions in common-pool resource management. Monitoring only group-average wealth can generate turn-taking, tolerate occasional errors, and sanction persistent overuse without identifying individual violators. The success of such governance depends strongly on the historical path by which institutions form and expand. These findings provide a mechanistic complement to institutional theory and highlight how simple decision principles can underlie complex forms of collective resource management.

\section*{Methods}
\subsection*{Model equations}
The algorithm for a game step in the model is as follows
\begin{align}
h_i(t) &= \chi(x(t) + S_i My_i(t) + A_i M\ev{y(t)}), \label{ds_eq:action} \\
r(x(t)) &= x(t) + \alpha(x(t) - x(t)^2), \label{ds_eq:growth} \\
x(t + 1)& = r(x(t))(1 - \beta \sum_{i=1}^{M} h_i(t)), \label{ds_eq:resource} \\
p_i(t) &= \beta h_i(t) r(x(t)), \label{ds_eq:harvest} \\
y_i(t + 1) &= \bigl(1 - \kappa\bigr)\,y_i(t) + p_i(t), \label{ds_eq:state}
\end{align}
where the threshold function $\chi$ is defined as $\chi(z)=1$ if $z>0$ and $\chi(z)=0$ otherwise. At each step, players decide their actions via Eq. \eqref{ds_eq:action} using the resource abundance, self wealth, and group-average wealth as inputs. Then, the resource grows according to Eq. \eqref{ds_eq:growth}, harvesting determines the remaining resource via Eq. \eqref{ds_eq:resource}, and wealth is updated via Eqs. \eqref{ds_eq:harvest} and \eqref{ds_eq:state}. Unless otherwise stated, we set \(\alpha = 1\), \(\beta = 0.9/M\), and \(\kappa = 0.25\).

\subsection*{Simulation protocol}
At the start of each generation, \(N\) groups of \(M\) players are formed and assigned fresh resource sites. The state variables are initialised as \(x(0)=0.1\) and \(y_i(0)=(1+\eta_i)/M\), where \(\eta_i\) is a small Gaussian noise with mean \(0\) and variance \(0.01\). Each generation consists of \(T=1000\) game steps. To exclude the initial transient, fitness is defined as the time-averaged wealth over the last $90\%$ of the game steps, $\bar y_i = \frac{1}{0.9T}\sum_{t=0.1T}^{T-1} y_i(t).$

After each generation, $NM$ offspring are produced. Each player is selected as a parent with probability proportional to its fitness. The resulting offspring counts approximately follow a Poisson distribution with the mean of fitness relative to the population-wide mean. Offspring inherit the parent's decision parameters \((S, A)\), with independent Gaussian mutations of mean \(0\) and variance \(\mu^2\) added to both parameters. The offspring are then randomly reassorted into groups in the next generation. Unless otherwise stated, we set \(\mu=0.1\), and each baseline fixed-$M$ simulation is run for 3000 generations.

\subsection*{Noise model}
To analyse stochasticity in governance, we model noise in the resource dynamics as
\[
r(x)=(2x-x^2)(1+\eta),
\]
and noise in peer monitoring as
\[
f_i(x(t),y_i(t),\ev{y(t)})=\chi\!\bigl(x(t)+S_i My_i(t)+A_i M\ev{y(t)}(1+\eta_i)\bigr),
\]
where \(\eta\sim \mathcal{N}(0,\sigma_r^2)\) and \(\eta_i\sim \mathcal{N}(0,\sigma_p^2)\).
The latter can also be interpreted as stochasticity in decision-making, because it is equivalent to introducing stochastic fluctuations in the effective decision threshold.

\section*{Data accessibility}
No empirical datasets were analysed in this study. The simulation outputs and source data underlying the figures, together with the code used for the simulations and analyses, are available at \url{https://github.com/KenjiItao/ds_game}.

\section*{Competing interests}
The authors declare no competing interests.

\newpage
\section*{Supplementary Text}
\supplementaryfigures
\supplementarytables
\subsection*{Robustness and limits of sustainable resource use}
The baseline within-generation dynamics are given by:
\begin{align}
h_i(t) &= \chi(x(t) + S_i My_i(t) + A_i M\ev{y(t)}), \label{ds_supp_eq:action} \\
r(x(t)) &= x(t) + \alpha(x(t) - x(t)^2), \label{ds_supp_eq:growth} \\
x(t + 1) &= r(x(t))\Bigl(1 - \beta \sum_{i=1}^{M} h_i(t)\Bigr), \label{ds_supp_eq:resource} \\
p_i(t) &= \beta h_i(t) r(x(t)), \label{ds_supp_eq:harvest} \\
y_i(t + 1) &= (1 - \kappa)\,y_i(t) + p_i(t), \label{ds_supp_eq:state}
\end{align}
where the threshold function $\chi$ is $\chi(z)=1$ for $z>0$ and $\chi(z)=0$ otherwise. At reproduction, offspring inherit $S$ and $A$ with independent Gaussian mutations of mean $0$ and variance $\mu^2$. Unless otherwise stated, $\alpha=1$, $\beta=0.9/M$, $\kappa=0.25$, and $\mu=0.1$. Figure S1 tests the sensitivity to each parameter. Sustainable resource use occurs across $0.5 \le \alpha \le 3.0$, $0.5 \le \beta M \le 1.2$, $0.1 \le \kappa \le 0.8$, and $0.01 \le \mu \le 10.0$. When $\beta M\ge1$, we cap the harvested fraction at $0.99$ to keep the resource level positive.

At the start of each generation, the resource is reset to $x(0)=0.1$ and individual wealth to $y_i(0)=(1+\eta_i)/M$, where $\eta_i$ is a Gaussian perturbation with mean $0$ and variance $0.01$. Scaling the initial wealth by $1/M$ (unlike the earlier two-player model \cite{itao2025self}) keeps it comparable to the long-run individual wealth across group sizes and limits the growth of the transient duration at large $M$. The results are almost unchanged when the initial resource abundance is set to $x(0)=0.3$ or $1.0$ (Fig. S15).

We use a linear decision rule for interpretability, but sustainable harvesting also evolves with recurrent neural-network (RNN) strategies, which allow nonlinear and history-dependent responses (Fig. S2). We further vary the social information available to each player: individual-specific wealth, group-average wealth, group-maximum wealth, or no peer information (Fig. S15).
While individual-specific monitoring and maximum wealth monitoring allow sustainability, without peer information, sustainability is largely restricted to small groups.

The threshold function $\chi$ makes each decision deterministic and abrupt. To test whether this is essential, we replaced it by a sigmoid while keeping the action binary, so that player $i$ harvests with probability $\sigma(k z_i)=1 / (1+\exp(-k z_i))$, where $z_i$ is the argument of $\chi$ in Eq.~(\ref{ds_supp_eq:action}) and $k$ sets the steepness. Because $z_i$ scales with the magnitude of the decision weights, we characterise the softness by the scale-free steepness $s = k\sqrt{S_i^2+A_i^2}$, which we either pin at $s=10$ or $s=100$, or let evolve through $\log k$ with the same mutation variance as $S$ and $A$ (Fig. S15). Sustainable resource use evolves in all three cases. A pinned, moderate steepness narrows the range of group sizes, whereas letting $s$ evolve widens it, because selection sharpens the decision rule where precision pays.

Gaussian mutations rarely produce large jumps in $(S,A)$. We therefore repeated the comparison with two heavy-tailed kernels: a symmetric $\alpha$-stable distribution with $\alpha=1.5$, and a symmetric Pareto tail with $P(|\Delta|>z)=z^{-a}$ and $a=2.5$ (Fig. S15). In both, the scale is chosen so that the median jump matches the Gaussian one and only the tail differs. Sustainable resource use still evolves over the baseline range of group sizes, and the evolved strategies occupy the same band, $S/A \simeq -1.1$, even with rare long-range mutations.

In the baseline model, fitness is the mean wealth over the last $90\%$ of the $T=1000$ game steps, the first $0.1T$ steps being discarded as the initial transient. Averaging over the full period from the outset often prevents sustainable strategies from evolving when $M$ is large. When overharvest occurs, players can accumulate wealth only in early steps before the resource collapses. Thus, counting these early gains makes it profitable to take as much as possible while the resource lasts. Discarding only the first $30$ steps also narrows the range supporting sustainability, though less severely. However, populations that first evolve under transient exclusion remain sustainable once selection switches to the full-period average (Fig. S15). Excluding the transient is thus required to escape overharvesting, but not to maintain the institution once it has emerged.

Random group formation is the baseline. To let offspring remain associated with their parental group, we introduce a migration rate $m$. Each position in a descendant group is filled from its founding group's offspring pool with probability $1-m$ and from the global offspring pool with probability $m$, so that smaller $m$ preserves stronger group structure and $m=1$ recovers the baseline. Smaller $m$ facilitates the evolution of sustainable resource use (Fig. S15), consistent with the stronger persistence of associations among strategies.

In the annealed runs of the main text, $M$ is multiplied by $\sqrt{2}$ every $1000$ generations ($M=2,3,4,6,\dots,128$). Two alternative schedules give the same picture: incrementing $M$ by one every $500$ generations ($M=2,\dots,100$), and slowing the $\sqrt{2}$ schedule to $3000$ generations per step (Fig. S15). Under all three, sustainable resource use persists to group sizes more than twice those reached at fixed $M$, and the strategies stay in the same band. Tripling the number of generations per step does not extend the range further, indicating that the limit is set by the group size itself rather than by the time available to adapt to each increase.

Ecological assumptions also matter. Keeping the individual harvesting fraction fixed independently of $M$ generally restricts sustainability, because the aggregate harvesting pressure then grows with $M$. With complete resource carry-over between generations, no sustainable outcome was observed. The present mechanism therefore requires resource renewal.

Figures S15 and S16 summarise these comparisons.
Sustainable resource use remains possible when fitness is defined by total harvest, decision rules are represented by RNNs, players can observe either the wealth of every group member or the group maximum, alternative schedules of group-size expansion are used, the threshold decision is replaced by a stochastic sigmoid, or mutations are heavy-tailed. Persistent group structure allows sustainability over a wider range of group sizes. By contrast, including the full initial transient, removing peer information, or keeping the individual harvesting fraction fixed generally restricts the range supporting sustainability.
A population that first evolved under transient exclusion can nevertheless maintain sustainability when fitness is subsequently averaged over all steps.
No sustainable outcome was observed when resources were carried over instead of being refreshed in each generation.

\subsection*{A long-run upper bound on total fitness}
To calculate the maximum fitness, we consider a harvesting game played by a single player. The dynamics of our $M$‐player game can be readily mapped onto this single‐player framework. For total group fitness, harvesting by $k$ players, each taking a fraction $\beta$, is equivalent to aggregate harvesting at intensity $k\beta$. A single aggregate resource user with a continuously adjustable harvesting intensity therefore provides an upper bound for the multiplayer game.

The dynamics of the single‐player game are given by
\begin{align}
    r(x(t)) &= (\alpha + 1)x(t) - \alpha x(t)^2,\\[1mm]
    x(t+1) &= r(x(t))(1 - \beta h(t)),\\[1mm]
    y(t+1) &= (1 - \kappa)y(t) + \beta h(t)r(x(t)) = (1 - \kappa)y(t) + p(t),
\end{align}
where \(p(t)\) represents the harvested amount at game step \(t\). To determine the maximum fitness, we assume that the resource growth rate \(\alpha\) and the state decay rate \(\kappa\) are given, and we optimise the harvesting fraction \(\beta\) to maximise \(\sum_t y(t)\).

Because \(r(x)\) is a concave function, it is optimal to harvest constantly at every game step (i.e., $h(t) \equiv 1$) rather than to accumulate the resource and harvest it in bulk over several periods. Accordingly, we focus on the steady state in which the player harvests at every step, so that
\[
x(t+1) \equiv x(t) \equiv x,\quad y(t+1) \equiv y(t) \equiv y,\quad \text{and} \quad p(t+1) \equiv p(t) \equiv p.
\]
In a steady state, the system reduces to
\begin{align}
    x &= \Bigl((\alpha + 1)x - \alpha x^2\Bigr)(1 - \beta), \label{eq:x}\\[1mm]
    p &= \frac{\beta}{1 - \beta}\,x,\label{eq:p}\\[1mm]
    y &= \frac{p}{\kappa}.
\end{align}
Solving \eqref{eq:x} for \(x\) yields
\begin{equation}
    x = \frac{\alpha - (\alpha + 1)\beta}{\alpha(1 - \beta)}.
\end{equation}
Substituting this expression into \eqref{eq:p} leads to the optimisation problem
\begin{equation}
    \max_\beta \; p = \frac{\beta\bigl(\alpha - (\alpha + 1)\beta\bigr)}{\alpha(1 - \beta)^2}.
\end{equation}
The optimal harvesting fraction \(\beta^\ast\) and the corresponding values of \(p^\ast\) and \(y^\ast\) are determined as follows:
\begin{align}
    p'(\beta) &= \frac{\alpha - (\alpha + 2)\beta}{\alpha(1-\beta)^3},\\[1mm]
    \beta^\ast &= \frac{\alpha}{\alpha + 2},\\[1mm]
    p^\ast &= \frac{\alpha}{4},\\[1mm]
    y^\ast &= \frac{\alpha}{4\kappa}.
\end{align}
Thus, the maximum fitness $\ev{y}^\ast$ is given by
\[
\ev{y}^\ast = y^\ast = \frac{\alpha}{4\kappa}.
\]

Hence, in the settings with $\alpha = 1$, $\kappa = 0.25$, and $\beta = 0.9/M$, the optimal harvesting frequency $F$ and the maximum total fitness $H$ are given by,
\begin{align}
    F &= \frac{\beta^\ast}{\beta M} = \frac{1}{2.7} \simeq 0.37 ,\\[1mm]
    H &= \frac{\alpha}{4\kappa} = 1.
\end{align}

\subsection*{Lyapunov stability of the within-generation dynamics}
Let us denote the state of a group as $\bm{z}(t)=\bigl(x(t),y_1(t),\ldots,y_M(t)\bigr)^{\mathsf T}$. Player $i$ harvests according to $h_i(t)=\chi\bigl(f_i(\bm{z}(t))\bigr)$ with the decision-making function $f_i(\bm{z})=x+S_iMy_i+A_iM\ev{y}$, and the dynamics are
\begin{align}
x(t+1)&=r(x(t))\Bigl(1-\beta\sum_{j=1}^{M}h_j(t)\Bigr),\label{eq:xdyn}\\[1mm]
y_i(t+1)&=(1-\kappa)y_i(t)+\beta h_i(t)r(x(t)),\label{eq:ydyn}
\end{align}
with $r(x)=x+\alpha (x-x^{2})$ and $r'(x)=(\alpha+1)-2\alpha x$.

The action $h_i$ is locally constant except on the switching boundary $f_i(\bm{z})=0$, so along a trajectory away from such boundaries $\partial h_i/\partial\bm{z}=\bm{0}$ and the Jacobian of \eqref{eq:xdyn} and \eqref{eq:ydyn} is
\begin{equation}
J(t)=
\begin{pmatrix}
\bigl(1-\beta\sum_j h_j(t)\bigr)r'(x(t)) & \bm{0}^{\mathsf T}\\[2mm]
\beta r'(x(t))\bm{h}(t) & (1-\kappa)I_M
\end{pmatrix},
\end{equation}
where $\bm{h}(t)=\bigl(h_1(t),\ldots,h_M(t)\bigr)^{\mathsf T}$. This matrix is block lower triangular, as are its products, so the Lyapunov spectrum splits into the contributions of the two diagonal blocks. The wealth block is the constant matrix $(1-\kappa)I_M$, giving
$\lambda_y=\log(1-\kappa)<0$ with multiplicity $M$, and the remaining exponent is
\begin{equation}
\lambda_x
=
\limsup_{T\rightarrow\infty}
\frac{1}{T}
\sum_{t=0}^{T-1}
\log\Bigl|\Bigl(1-\beta\textstyle\sum_j h_j(t)\Bigr)r'(x(t))\Bigr|.
\label{eq:lamx}
\end{equation}

\eqref{eq:xdyn} gives $1-\beta\sum_j h_j(t)=x(t+1)/r(x(t))$, which removes the harvesting factor from \eqref{eq:lamx} and leaves a telescoping product,
\begin{equation}
\prod_{t=0}^{T-1}
\Bigl|\Bigl(1-\beta\textstyle\sum_j h_j(t)\Bigr)r'(x(t))\Bigr|
=
\frac{x(T)}{x(0)}
\prod_{t=0}^{T-1}
\left|\frac{(\alpha+1)-2\alpha x(t)}{(\alpha+1)-\alpha x(t)}\right|.
\label{eq:telescope}
\end{equation}
Thus, the harvesting history enters $\lambda_x$ only through the resource levels visited. For $0<x(t)<(\alpha+1)/\alpha$, each factor of the second product is bounded by $1$ if and only if
\begin{equation}
x(t)\leq\frac{2(\alpha+1)}{3\alpha}.
\label{eq:ratiobound}
\end{equation}
Because $0<1-\beta M\leq1-\beta\sum_j h_j(t)\leq1$ for $\beta M<1$, the resource satisfies
\begin{equation}
0<x(t+1)\leq r(x(t))\leq\frac{(\alpha+1)^2}{4\alpha}.
\end{equation}
Provided that $x(0)$ satisfies \eqref{eq:ratiobound}, as it does in the model, this bound implies by induction that \eqref{eq:ratiobound} holds at every step if
\begin{equation}
\frac{(\alpha+1)^2}{4\alpha}\leq\frac{2(\alpha+1)}{3\alpha},
\qquad\text{that is,}\qquad
\alpha\leq\frac{5}{3}.
\label{eq:alphacond}
\end{equation}
This condition also keeps $[0,(\alpha+1)/\alpha]$ forward invariant. Since $x(T)$ is bounded above, the factor $x(T)/x(0)$ in \eqref{eq:telescope} cannot make a positive asymptotic contribution. Consequently,
\begin{equation}
\lambda_{\max}=\max\left\{\lambda_x,\log(1-\kappa)\right\}\leq0
\end{equation}
independently of the strategy parameters and of the realised harvesting pattern.

Hence, in the settings with $\alpha=1$, $\kappa=0.25$, and $\beta=0.9/M$, we have $r'(x)=2(1-x)$ and $0<x(t)\leq1$, whereas the bound \eqref{eq:ratiobound} permits $x(t)\leq4/3$. Condition (\ref{eq:alphacond}) is therefore met with a comfortable margin, and $\lambda_y=\log(0.75)\simeq-0.288$. Moreover, for a periodic mode of period $P$, $x(t+P)=x(t)$ and the telescoping term vanishes, giving
\begin{equation}
\lambda_x
=\frac{1}{P}\sum_{t=0}^{P-1}
\log\left(\frac{2(1-x(t))}{2-x(t)}\right)<0.
\label{eq:periodic_lamx}
\end{equation}
Thus, every periodic attractor observed under the baseline parameters has $\lambda_{\max}<0$. If severe overharvesting instead causes $x(t)\to0$, ~\eqref{eq:telescope} still gives $\lambda_x\leq0$, with equality possible only at a neutral boundary. The baseline parameter values therefore preclude chaotic dynamics characterised by a positive Lyapunov exponent.

Condition~(\ref{eq:alphacond}) is sufficient but not necessary and directly covers $0.5\leq\alpha\leq5/3$ in the parameter range of Fig.~S1. For a constant harvesting factor $1-\beta\sum_j h_j\equiv c$, the change of variable $u=\alpha x/(\alpha+1)$ reduces \eqref{eq:xdyn} to the logistic map $u(t+1)=c(\alpha+1)u(t)(1-u(t))$. At the optimal intensity $\beta^\ast=\alpha/(\alpha+2)$, one has $(1-\beta^\ast)(\alpha+1)=2(\alpha+1)/(\alpha+2)<2$ and the steady state is $x=1/2$ for every $\alpha$, showing that efficient constant harvesting strongly damps the resource dynamics. This observation does not, however, prove the absence of chaos for $\alpha>5/3$, because the realised factor $c(t)=1-\beta\sum_jh_j(t)$ generally varies over time. Indeed, at $\alpha=3$ the no-harvesting branch $c=1$ is the logistic map with parameter $4$. Stability over the remainder of the range in Fig.~S1 must therefore be assessed from the realised trajectories using ~\eqref{eq:lamx}, rather than inferred from the optimal constant-harvesting state.

This calculation is based on the fact that $d\chi/d\bm{z}=\bm{0}$, except on the boundary $f_i(\bm{z})=0$ at which $\chi$ jumps between $0$ and $1$. Modes whose orbit visits this boundary are an exception to the above calculation. However, we ignore them because such $\bm{z}$ form a set of measure zero. For a periodic mode satisfying $\min_{i,t}|f_i(\bm{z}(t))|>0$, a neighbourhood of the orbit has the same action itinerary, and the negative exponent in \eqref{eq:periodic_lamx} establishes its local asymptotic stability. In our simulations, the asymptotic trajectories converged to stable limit cycles or to the resource-collapse state, and we found no evidence of trajectories exhibiting persistent boundary-induced irregularity.

\clearpage

\section*{Supplementary Figures}
\begin{figure}[!htbp]
  \centering
   \includegraphics[width=\linewidth]{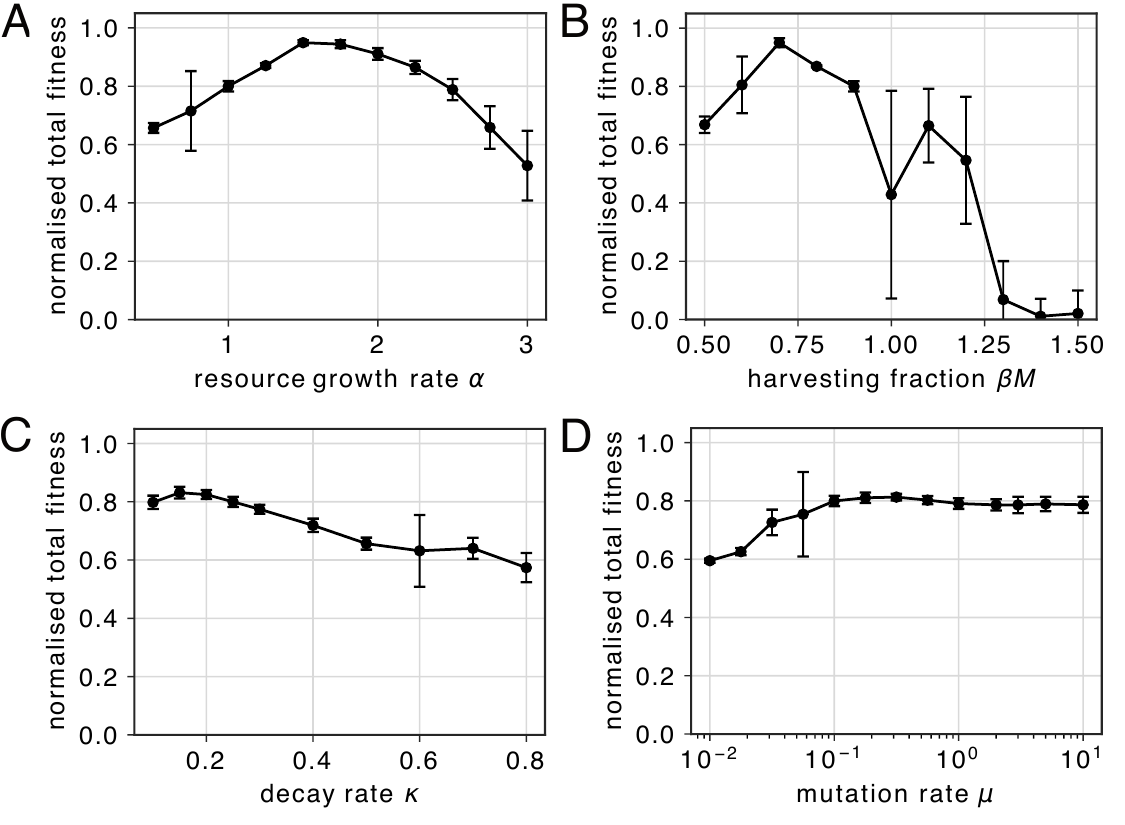}
   \caption{Robustness of simulation results to fixed parameters.
Dependence of the mean total fitness (normalised by the theoretical maximum, $\alpha / 4\kappa$ for each parameter condition) on (A) resource growth rate $\alpha$, (B) harvesting parameter $\beta M$, (C) state‐decay rate $\kappa$, and (D) mutation rate $\mu$.
Points show means over 100 trials; error bars denote one standard deviation.
Unless varied on the $x$‐axis, parameters are held at $\alpha=1$, $\beta M=0.9$, $\kappa=0.25$, $M=5$, $N=100$, and $\mu=0.1$. Sustainable resource governance consistently emerged across $0.5 \le \alpha \le 3.0$, $0.5 \le \beta M \le 1.2$, $0.1 \le \kappa \le 0.8$, and $0.01 \le \mu \le 10.0$. When \(\beta M \ge 1\), to avoid negative resource values, we replace the update rule with
$x(t+1)=r\bigl(x(t)\bigr)\max\Bigl(0.01,\bigl(1-\beta\sum_{i=1}^{M}h_i(t)\bigr)\Bigr). $ Then, players' payoff is replaced with $
p_i(t)=h_i(t)r\bigl(x(t)\bigr)\min\Bigl(0.99,\beta\sum_{i=1}^{M}h_i(t)\Bigr) / \sum_{i=1}^{M}h_i(t),$
when \(\sum_i h_i(t)>0\), and \(p_i(t)=0\) when \(\sum_j h_j(t)=0\).}
\label{fig:DS_game_Mplayer_robustness}
\end{figure}

\begin{figure}[!htbp]
  \centering
   \includegraphics[width=\linewidth]{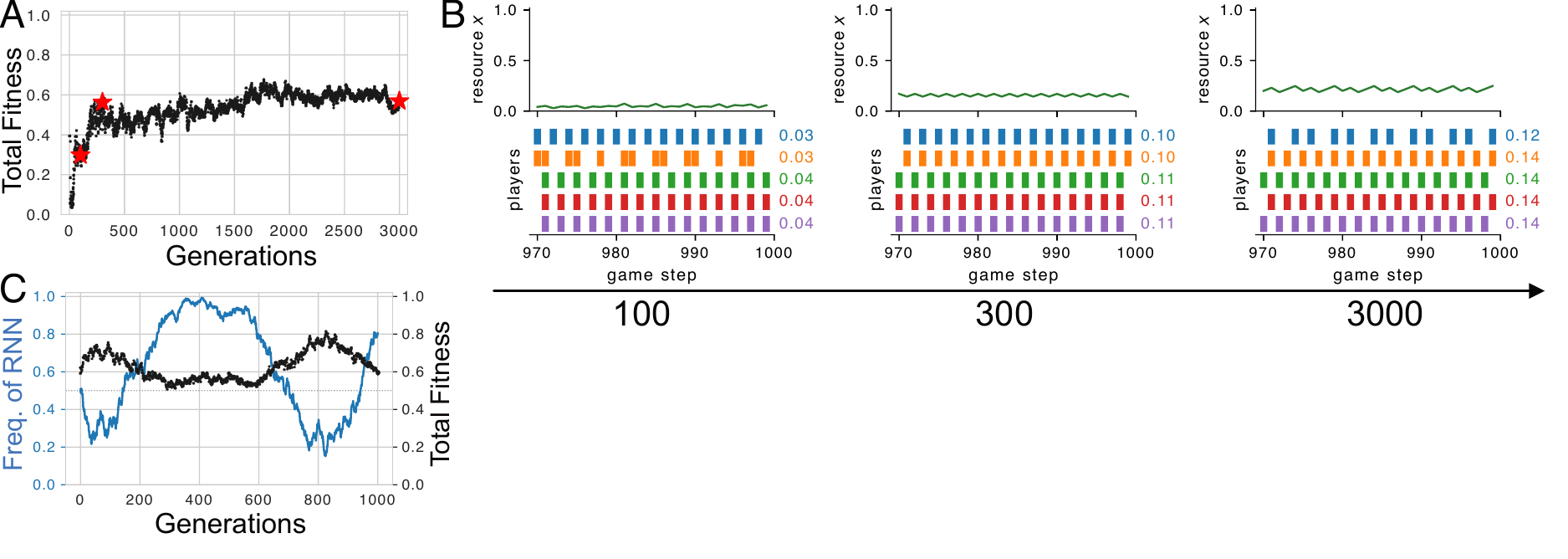}
   \caption{Robustness of the simulation results when strategies are implemented as recurrent neural networks (RNNs) with $4$ hidden nodes, taking the resource abundance $x(t)$, self wealth $My_i(t)$, and group-average wealth $M\ev{y(t)}$ as inputs (results are qualitatively unchanged with $8$ hidden nodes). Shown is an example of the evolutionary dynamical-systems game with $M=5$ players and $N=100$ groups.
(A) Changes in the mean total fitness across the $N$ groups over generations.
(B) Game dynamics during the final 30 steps (steps $970$--$999$). Green curves show the resource dynamics. Coloured squares and values indicate the harvesting actions and fitness of players, respectively.
Emergent patterns are similar to those in the main text obtained with linear decision rules.
(C) Mixed evolutionary simulation of perceptron and RNN strategies. We first evolved perceptron-based and RNN-based populations independently for $3000$ generations, then mixed the evolved strategies and continued evolution for an additional $1000$ generations. The blue line (left axis) indicates the frequency of RNN strategies in the mixed population, while black dots (right axis) show the mean total fitness across the $N$ groups.
Across runs, RNN strategies took over the population in some cases, perceptron strategies took over in others, and the two coexisted in the rest. The run shown here is an example of coexistence.
}
\label{fig:DS_game_Mplayer_RNN}
\end{figure}

\begin{figure}[!htbp]
  \centering
   \includegraphics[width=\linewidth]{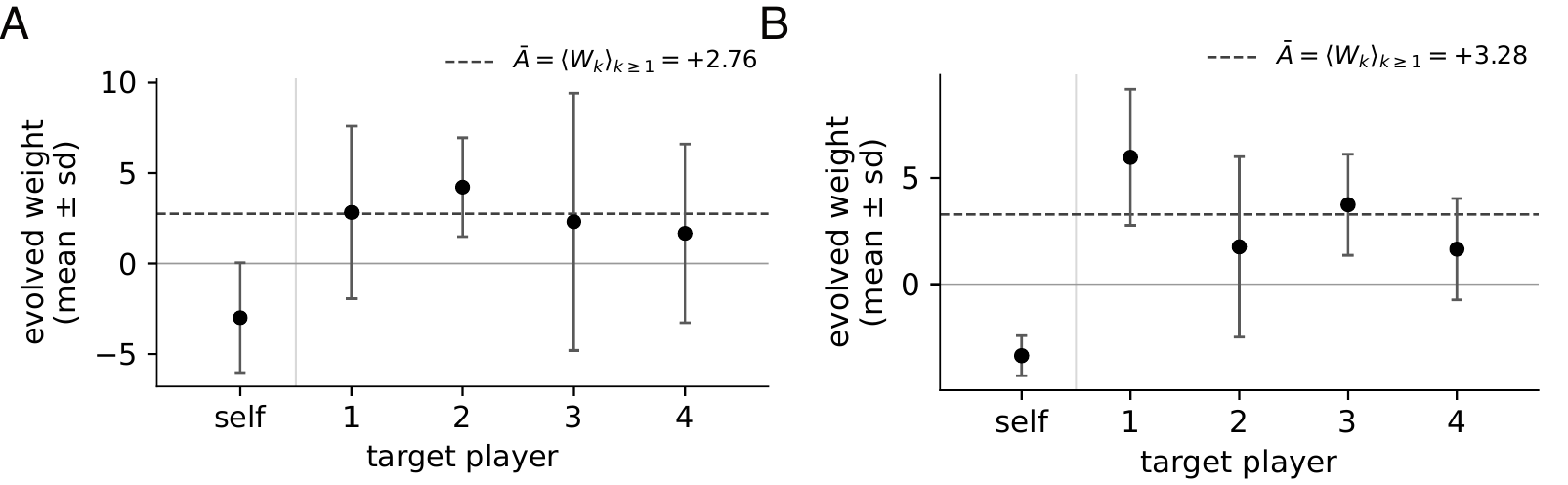}
   \caption{Emergent simplicity in decision rules with individual-specific monitoring.
   Each player carries $M$ evolving weights: one ($W_0$) for its own wealth and $M-1$ ($\{W_k\}_{k = 1}^{M-1}$) for co-players' wealth.
(A) The remaining players are indexed by their relative positions, $k=j-i \pmod M$, and harvesting occurs when $x+W_0My_i+M^{-1}\sum_{k=1}^{M-1}W_kMy_{i+k}>0$.
(B) The other players are instead ordered by wealth, $y_{(1)}\geq y_{(2)}\geq\cdots\geq y_{(M-1)}$, with $y_{(1)}$ denoting the wealthiest player other than the focal player, and harvesting occurs when $x+W_0My_i+M^{-1}\sum_{k=1}^{M-1}W_kMy_{(k)}>0$.
Points and error bars show the mean and standard deviation of the evolved weights after $3000$ generations for $N=100$ and $M=5$, pooled across all individuals in five sustainable trials (i.e., with total fitness $> 0.3$).
The dashed line indicates the mean weight assigned to other players.
In both models, the mean self-weight $W_0$ evolves to be negative, whereas the weights assigned to other players are positive and remain centred around a common value $\bar A$. Thus, despite the availability of $M$ independent weights, the evolved rules approximately collapse to two effective components: a negative self-weight and a common positive weight for peer wealth, corresponding to the simple principle \emph{wait when wealthy, harvest when below average}.
}
\label{fig:DS_game_Mplayer_simplicity}
\end{figure}

\begin{figure}[!htbp]
  \centering
   \includegraphics[width=\linewidth]{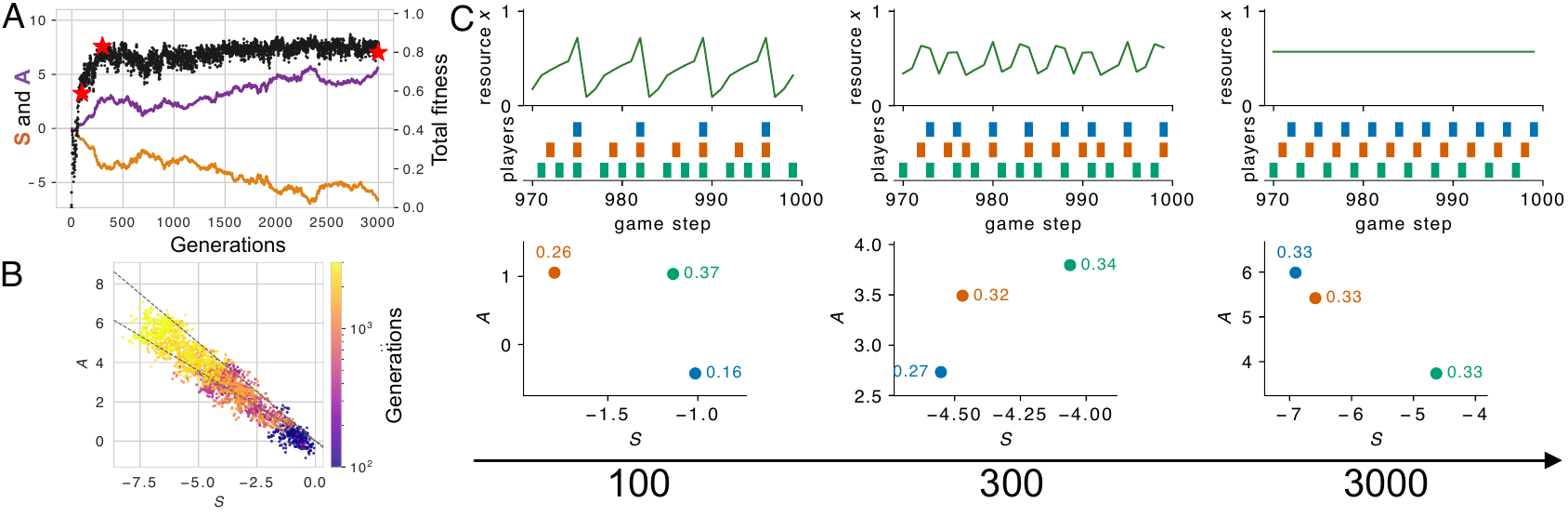}
   \caption{Example of the evolutionary dynamical-systems game with $M = 3$ players and $N = 100$ groups.
(A) Evolution of the population averages of the decision parameters $S$ (orange) and $A$ (purple). Black dots denote the mean total fitness of $N$ groups (right axis). Red stars mark the generations highlighted in (C).
(B) Temporal change in the distribution of decision parameters. Dashed lines show $S/A = -1.1$ and $-1.3$.
(C) Game dynamics during the final 30 steps (steps $970$–$999$). Green curves show resource abundance (top), and coloured squares mark each player's harvesting actions (middle). The bottom panels show each player's $(S,A)$ values and fitness in the corresponding colour. Numbers beneath each panel give the generation index.}
    \label{fig:DS_game_Mplayer_temporal_M3}
\end{figure}

\begin{figure}[!htbp]
  \centering
   \includegraphics[width=\linewidth]{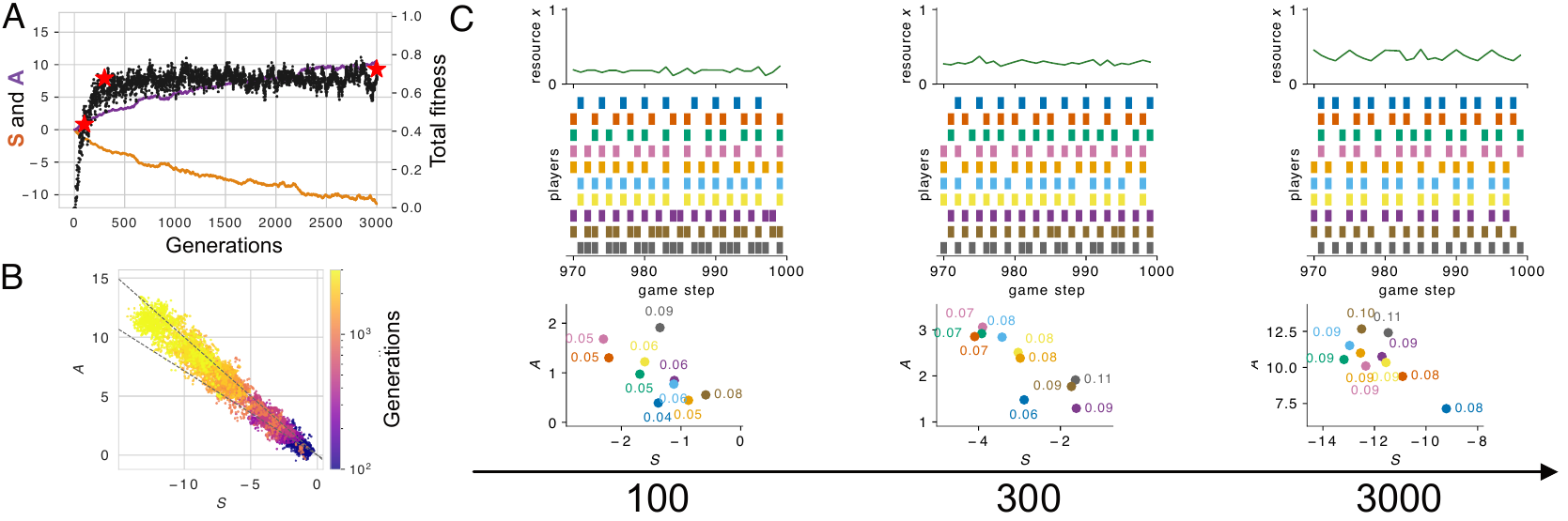}
      \caption{Example of the evolutionary dynamical-systems game with $M = 10$ players and $N = 100$ groups.
(A) Evolution of the population averages of the decision parameters $S$ (orange) and $A$ (purple). Black dots denote the mean total fitness of $N$ groups (right axis). Red stars mark the generations highlighted in (C).
(B) Temporal change in the distribution of decision parameters. Dashed lines show $S/A = -1.1$ and $-1.3$.
(C) Game dynamics during the final 30 steps (steps $970$–$999$). Green curves show resource abundance (top), and coloured squares mark each player's harvesting actions (middle). The bottom panels show each player's $(S,A)$ values and fitness in the corresponding colour. Numbers beneath each panel give the generation index.}
    \label{fig:DS_game_Mplayer_temporal_M10}
\end{figure}

\begin{figure}[!htbp]
  \centering
   \includegraphics[width=\linewidth]{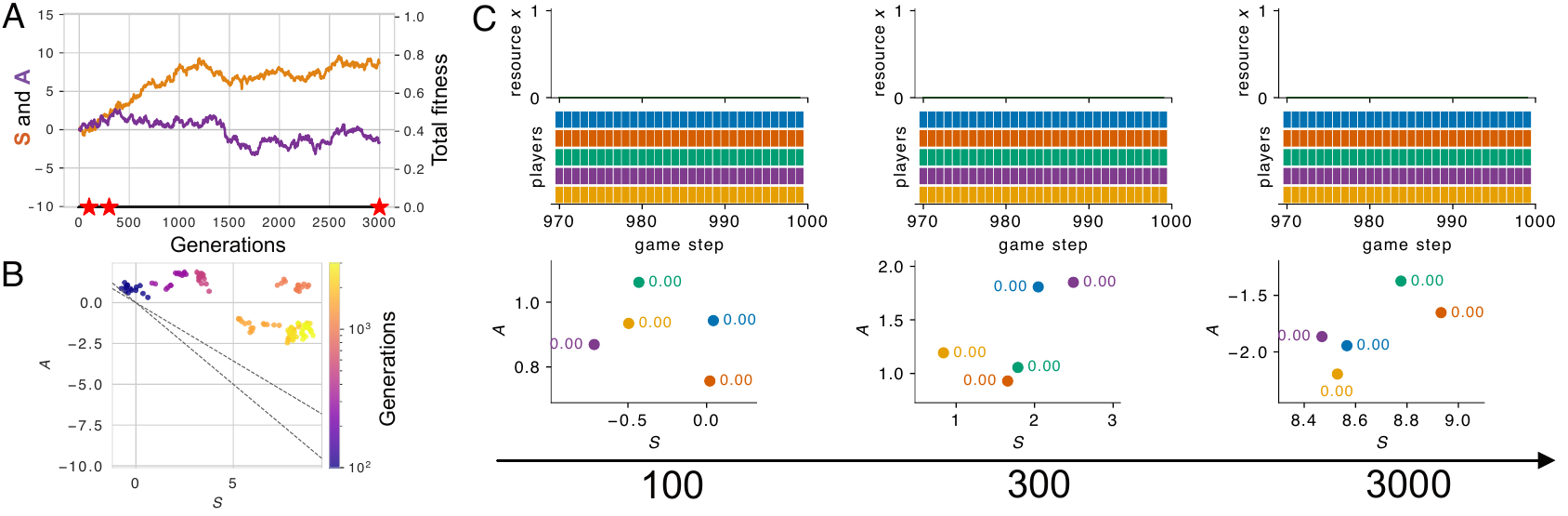}
   \caption{Example of the evolutionary dynamical-systems game with $M = 5$ players and $N = 3$ groups (i.e., the condition where sustainability does not evolve).
(A) Evolution of the population averages of the decision parameters $S$ (orange) and $A$ (purple). Black dots denote the mean total fitness of $N$ groups (right axis). Red stars mark the generations highlighted in (C).
(B) Temporal change in the distribution of decision parameters. Dashed lines show $S/A = -1.1$ and $-1.3$.
(C) Game dynamics during the final 30 steps (steps $970$–$999$). Green curves show resource abundance (top), and coloured squares mark each player's harvesting actions (middle). The bottom panels show each player's $(S,A)$ values and fitness in the corresponding colour. Numbers beneath each panel give the generation index.}
    \label{fig:DS_game_Mplayer_temporal_M5_N3}
\end{figure}

\begin{figure}[!htbp]
  \centering
   \includegraphics[width=\linewidth]{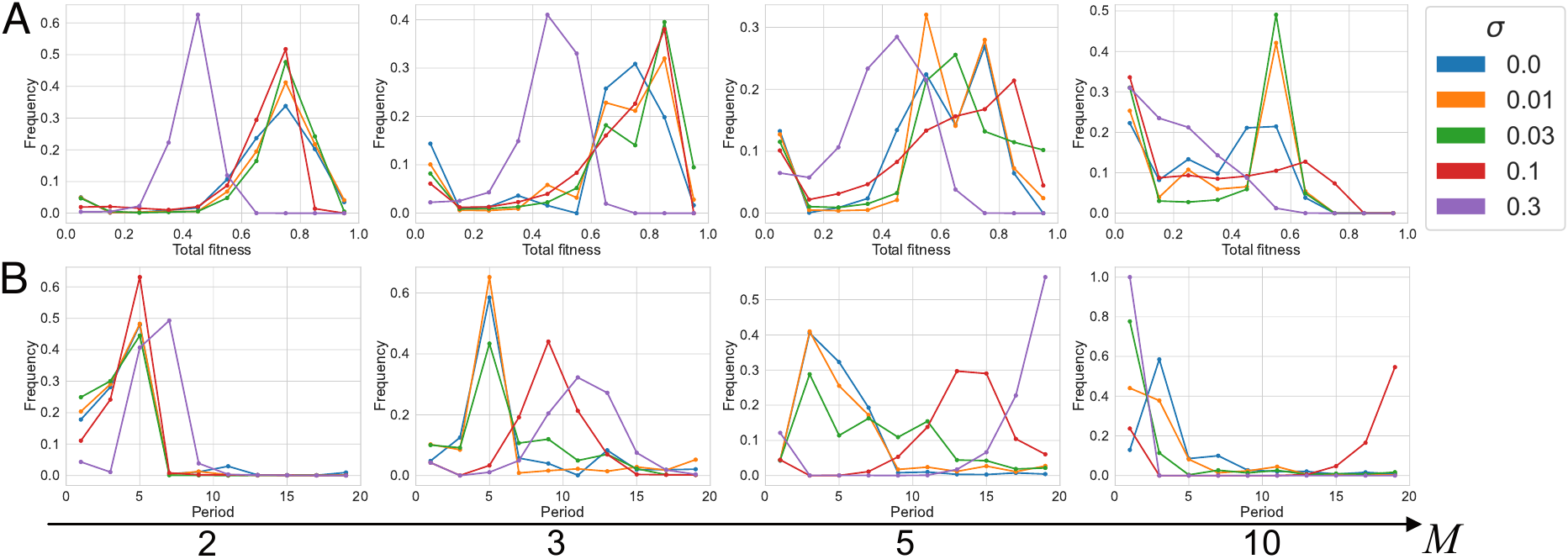}
    \caption{Dependence of total fitness (A) and periods of modes (B) on strategic diversity $\sigma$. In this analysis, $M$ strategies are randomly generated around each $(S, A)$, using independent multiplicative Gaussian perturbations with mean $1$ and variance $\sigma^2$. $(S, A)$ are uniformly sampled from the cooperative regions, given by $-25 \le S \le 0$ and $-1.3 \le S/A \le -1.1$. When $M \ge 3$, the total fitness is maximised under moderate $\sigma$, and the period generally increases with $\sigma$.}
    \label{fig:DS_game_Mplayer_period_sigma_si}
\end{figure}

\begin{figure}[!htbp]
  \centering
   \includegraphics[width=\linewidth]{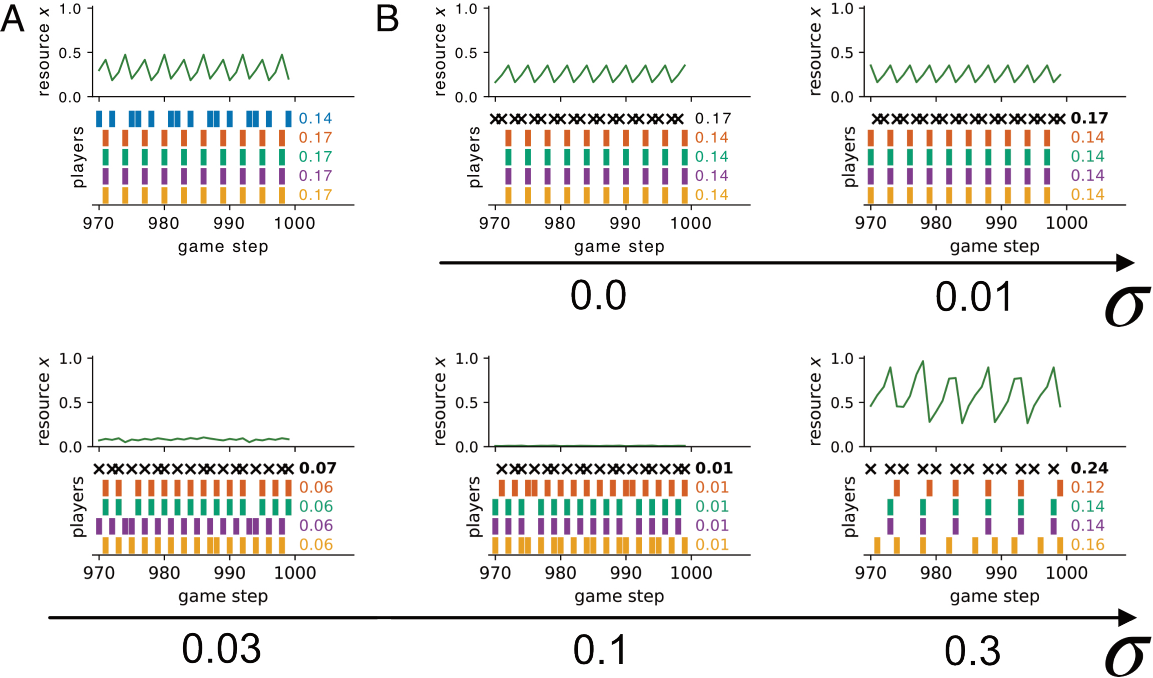}
   \caption{Responses against mutants' invasion. Four resident strategies (coloured boxes) are drawn from a distribution centred at $(S, A)=(-4, 3.4)$, using independent multiplicative Gaussian perturbations with mean $1$ and variance $\sigma^2$. By contrast, player 1 (the selfish mutant, plotted as $\times$) adopts strategy $(S, A)=(-3, 3.4)$, who harvests more frequently than residents.
(A) Baseline behaviour of a homogeneous group of five residents with $(S, A)=(-4, 3.4)$.
(B) Resident responses to mutant invasion as a function of strategic diversity $\sigma$. When $\sigma$ is small, players cluster into two—mutant versus residents—permitting the mutant to exploit residents. Conversely, at intermediate $\sigma$, some residents increase their harvesting frequency to punish the mutant, thereby reducing the mutant’s fitness. Although, under intermediate $\sigma$, the mutant's fitness is larger than that of the residents, its fitness is smaller than the original fitness in (A), achieved when the mutant follows the residents' strategy. Hence, with the current punishment mechanisms, the evolutionary robustness of the residents is maintained under intermediate $\sigma$.
}
    \label{fig:DS_game_Mplayer_er_mode}
\end{figure}

\begin{figure}[!htbp]
  \centering
   \includegraphics[width=\linewidth]{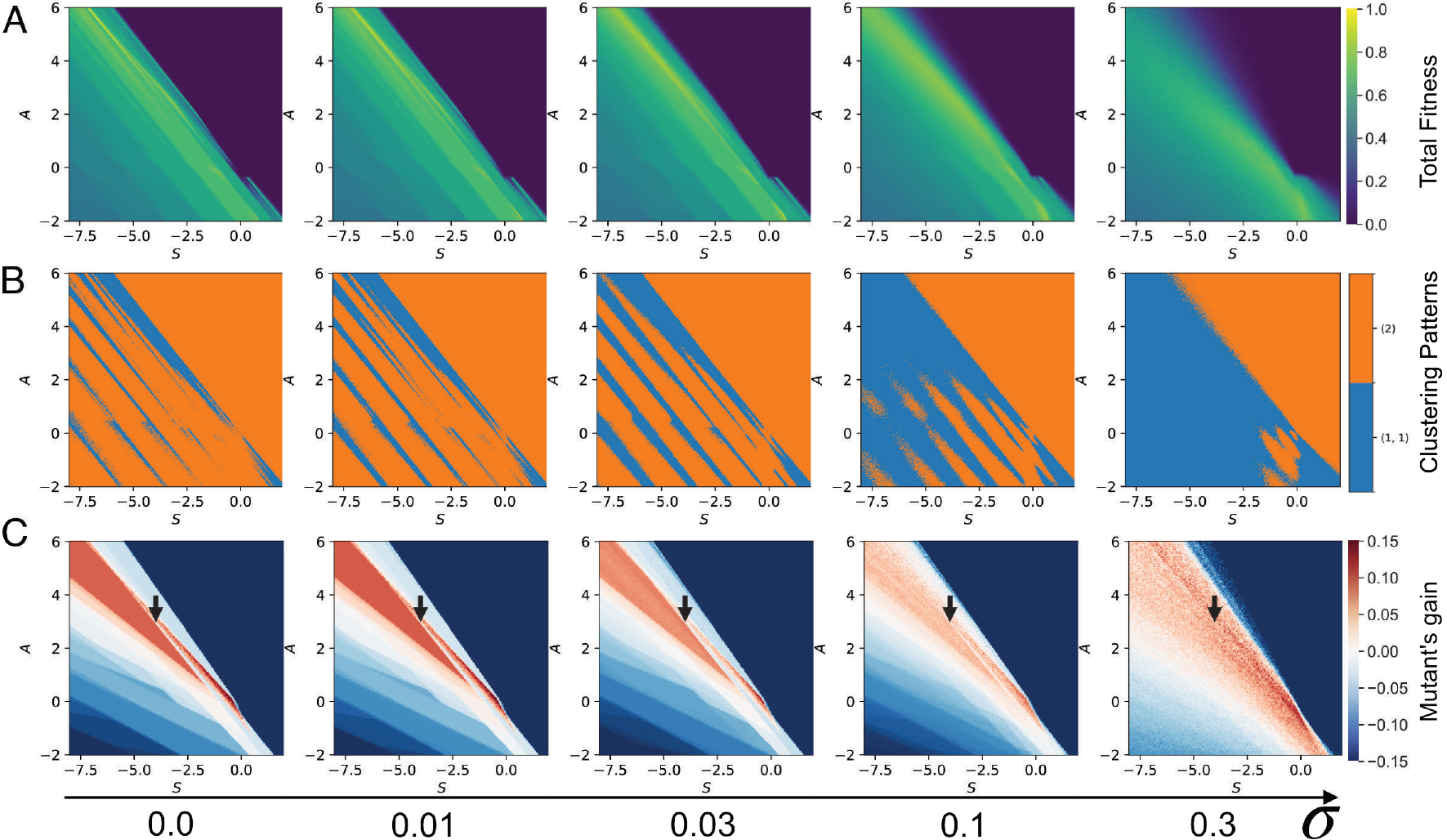}
   \caption{Decision parameter dependence of fitness and clustering patterns for varying strategic variance with $M = 2$ players. (A) The total fitness of players in the group. (B) Clustering patterns of players' action sequences; for example, ``(2)'' denotes synchrony, and ``(1, 1)'' denotes the labour division.
   In (A) and (B), two strategies are generated by multiplying Gaussian perturbations with mean $1$ and variance \(\sigma^{2}\) to each \((S, A)\).
   (C) Evolutionary robustness. A strategy is fixed around the arrowhead at $(S, A)=(-4, 3.3)$, while the mutant's strategy is varied in parameter space. Colours indicate the difference between the mutant's fitness and the average fitness when all strategies are around the arrowhead. A fixed strategy can be invaded by those plotted in red. Panels show mean total fitness, the most frequent clustering pattern, and mean mutant fitness gain across $100$ trials for each $(S,A)$. The benefits of strategic diversity remain minimal when $M = 2$.}
    \label{fig:DS_game_Mplayer_phase_M2}
\end{figure}

\begin{figure}[!htbp]
  \centering
   \includegraphics[width=\linewidth]{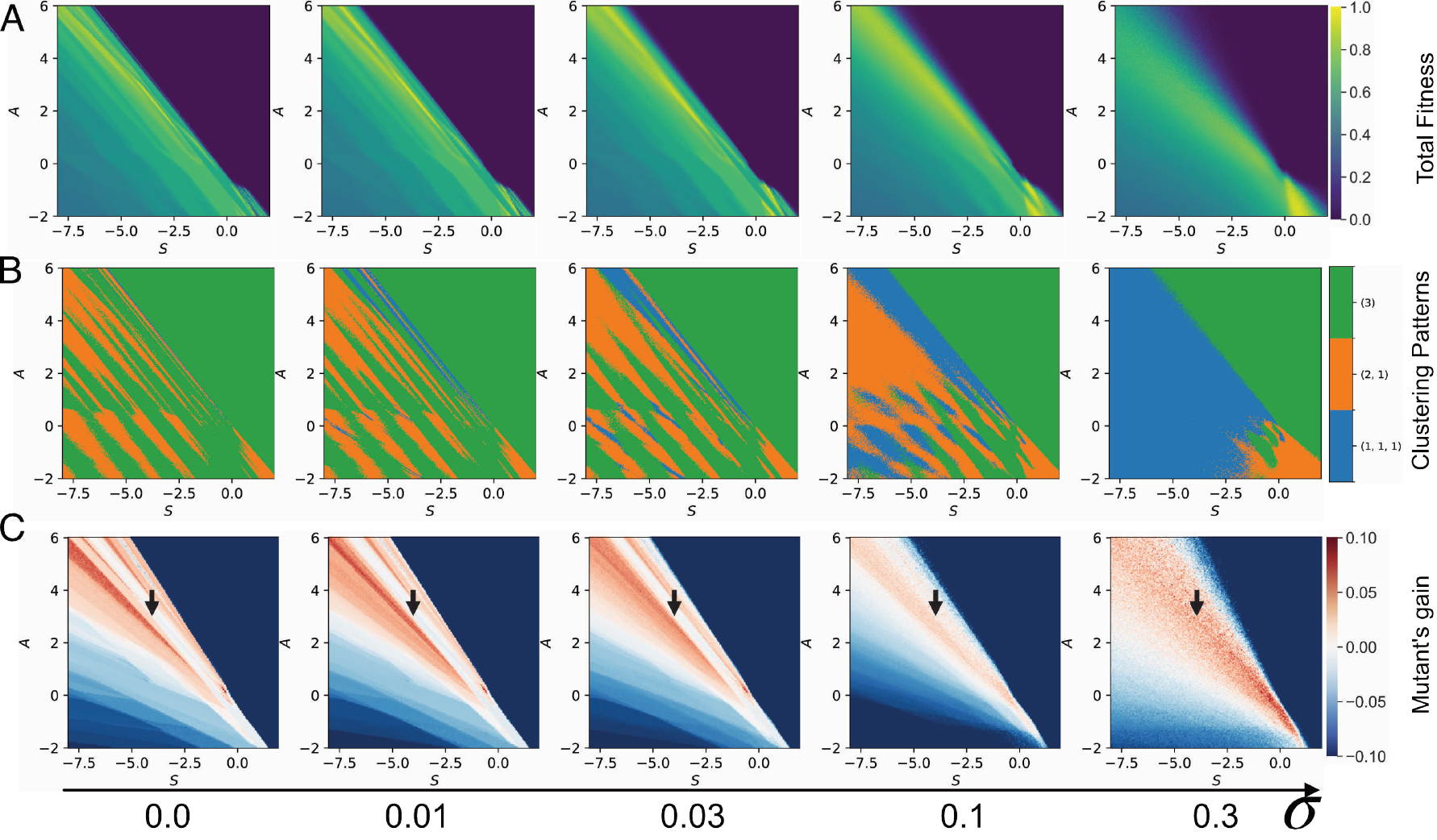}
   \caption{Decision parameter dependence of fitness and clustering patterns for varying strategic variance with $M = 3$ players. (A) The total fitness of players in the group. (B) Clustering patterns of players' action sequences; for example, ``(2, 1)'' denotes two clusters with two and one players.
   In (A) and (B), three strategies are generated by multiplying Gaussian perturbations with mean $1$ and variance \(\sigma^{2}\) to each \((S, A)\).
   (C) Evolutionary robustness. Two strategies are fixed around the arrowhead at $(S, A)=(-4, 3.3)$, while the mutant's strategy is varied in parameter space. Colours indicate the difference between the mutant's fitness and the average fitness when all strategies are around the arrowhead. Fixed strategies can be invaded by those plotted in red. Panels show mean total fitness, the most frequent clustering pattern, and mean mutant fitness gain across $100$ trials for each $(S,A)$. When $M = 3$, the total fitness is maximised under $\sigma = 0.03$, and the evolutionary robustness is maximised under $\sigma = 0.1.$}
    \label{fig:DS_game_Mplayer_phase_M3}
\end{figure}

\begin{figure}[!htbp]
  \centering
   \includegraphics[width=\linewidth]{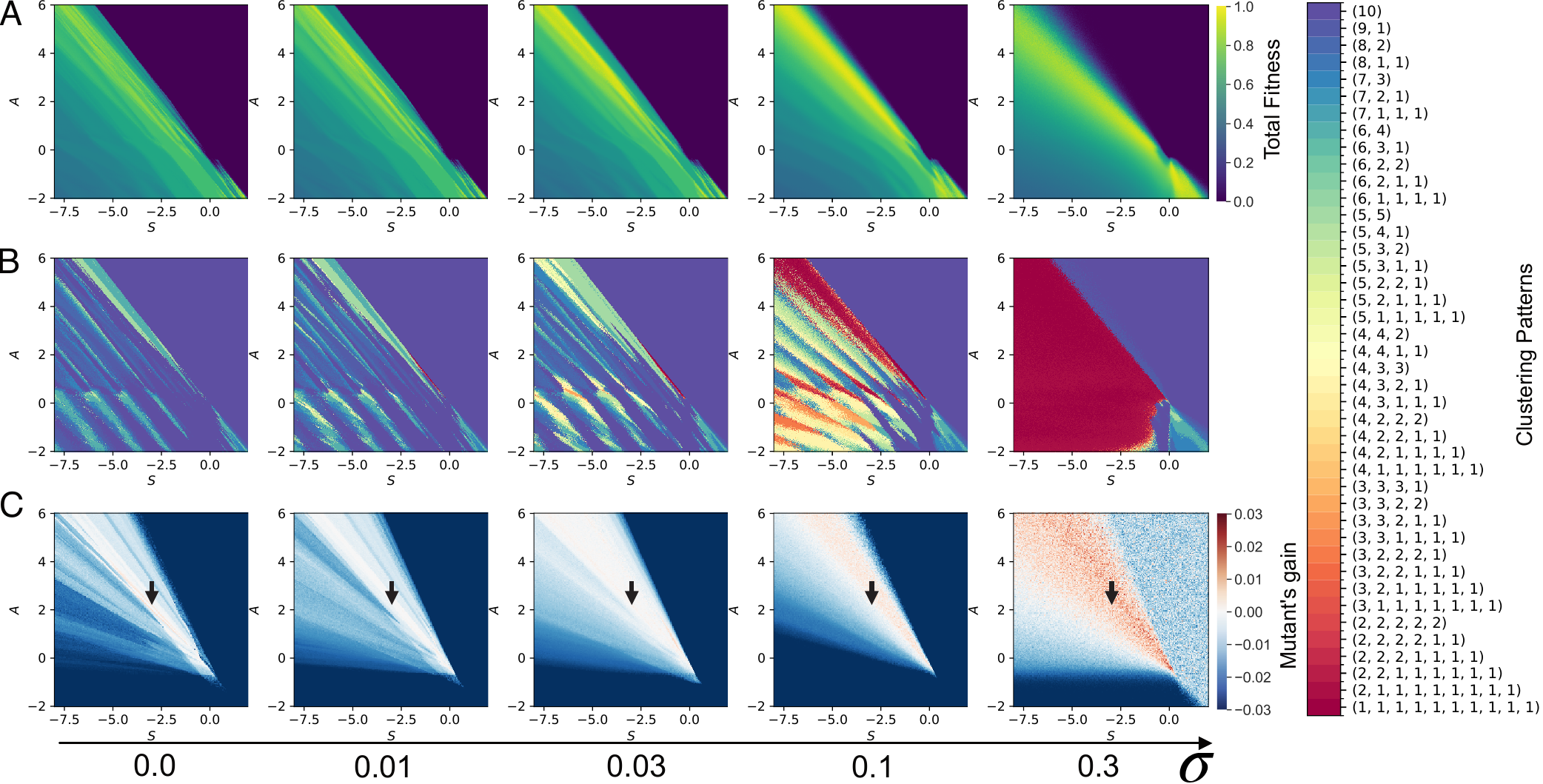}
   \caption{Decision parameter dependence of fitness and clustering patterns for varying strategic variance with $M = 10$ players. (A) The total fitness of players in the group. (B) Clustering patterns of players' action sequences; for example, ``(7, 3)'' denotes two clusters with seven and three players. In (A) and (B), ten strategies are generated by multiplying Gaussian perturbations with mean $1$ and variance \(\sigma^{2}\) to each \((S, A)\).
   (C) Evolutionary robustness. Nine strategies are fixed around the arrowhead at $(S, A)=(-3, 2.5)$, while the mutant's strategy is varied in parameter space. Colours indicate the difference between the mutant's fitness and the average fitness when all strategies are around the arrowhead. Fixed strategies can be invaded by those plotted in red. Panels show mean total fitness, the most frequent clustering pattern, and mean mutant fitness gain across $100$ trials for each $(S,A)$. When $M = 10$, strategic diversity greatly increases total fitness.}
    \label{fig:DS_game_Mplayer_phase_M10}
\end{figure}

\begin{figure}[!htbp]
  \centering
   \includegraphics[width=\linewidth]{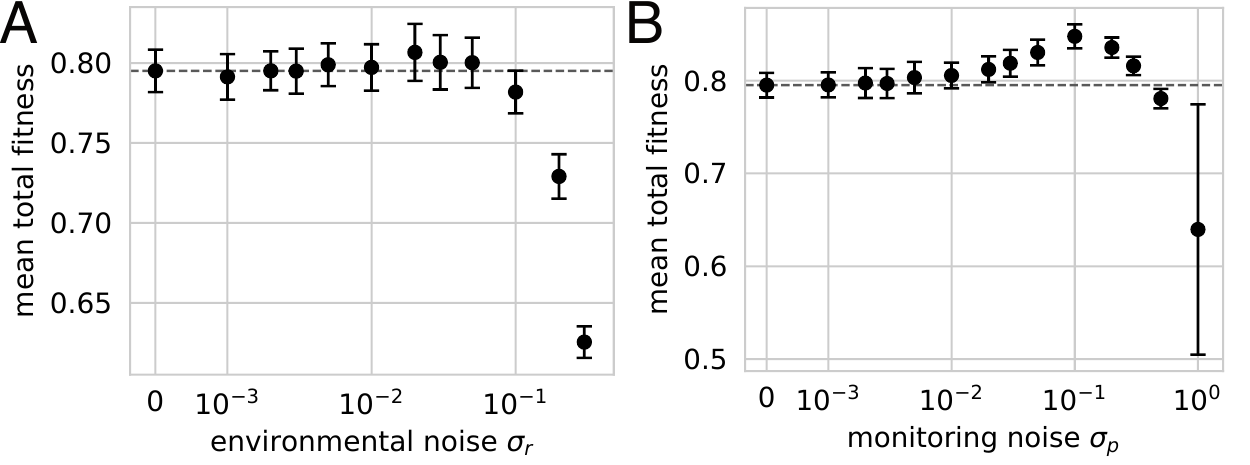}
    \caption{Dependence of the total fitness on noise levels in the resource dynamics (A) and peer monitoring (B). Dots show the mean total fitness, and error bars indicate the standard deviation over $100$ trials for $N = 100$ and $M=5$.
    (A) Environmental noise. The resource dynamics follow $r(x)=(2x-x^{2})(1+\eta)$, where $\eta\sim\mathcal{N}(0,\sigma_r^2)$ is a Gaussian perturbation. Fitness is maximised at $\sigma_r=0.02$ and drops sharply for $\sigma_r\ge 0.2$.
    (B) Monitoring noise. Player $i$'s decision-making follows $h_i(t)=f_i(x(t), y_i(t),\ev{y(t)})=\chi\!\bigl(x(t)+S_i My_i(t)+A_i M\ev{y(t)}(1+\eta_i)\bigr)$, where $\eta_i\sim\mathcal{N}(0,\sigma_p^2)$ is a Gaussian perturbation. Fitness is maximised at $\sigma_p=0.1$ and drops sharply for $\sigma_p\ge 0.5$. Note that the monitoring noise can also be interpreted as stochasticity in decision-making, because it is equivalent to introducing stochastic fluctuations in the effective decision threshold.}
    \label{fig:DS_game_Mplayer_noise}
\end{figure}

\begin{figure}[!htbp]
  \centering
   \includegraphics[width=\linewidth]{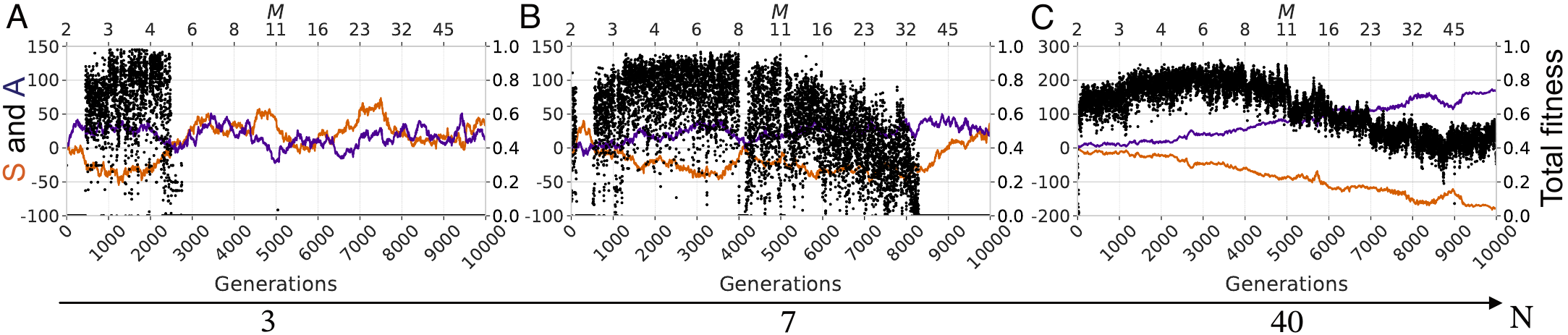}
   \caption{Evolutionary simulation with gradual expansion of group size $M$.
Evolution of the population‐averaged decision parameters $S$ (orange) and $A$ (purple) is shown; black dots mark the total mean fitness (right axis). The numbers along the top indicate the current value of $M$, which is multiplied by $\sqrt{2}$ every 1000 generations.
Although total fitness declines and overharvesting eventually emerges—marked by a rise in $S$ that drives selfish resource use—sustainability persists for large $M$ that would trigger collapse under a fixed-$M$ scenario. Numbers beneath each panel give the number of groups $N$.
}
    \label{fig:DS_game_Mplayer_anneal}
\end{figure}

\begin{figure}[!htbp]
  \centering
   \includegraphics[width=\linewidth]{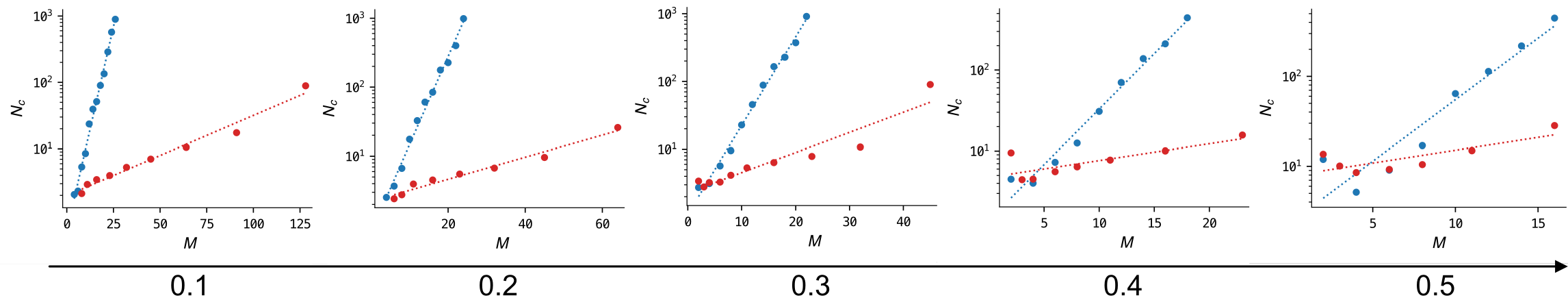}
   \caption{Robustness of phase-boundary scaling to the sustainability threshold. Each panel shows the minimum number of groups required for sustainability, $N_c$, as a function of group size $M$. From left to right, sustainability is defined as total group fitness exceeding $0.1$, $0.2$, $0.3$, $0.4$, or $0.5$, while the main text uses $0.3$.
   Blue and red indicate fixed group size and gradual group-size expansion, respectively. Points show the simulated phase boundaries, and dashed lines show exponential fits, $N_c\propto\exp(aM)$. For thresholds of $0.1$, $0.2$, $0.3$, $0.4$, and $0.5$, respectively, the fitted exponents $a$ are $0.28$, $0.29$, $0.30$, $0.32$, and $0.32$ for fixed $M$, and $0.03$, $0.04$, $0.07$, $0.05$, and $0.07$ for gradual expansion. Exponential fits consistently outperform power-law fits in all conditions.
   }
\label{fig:DS_game_Mplayer_robustness_N_qc}
\end{figure}

\clearpage

\begin{center}
  \captionsetup{type=figure}

  \includegraphics[
    width=\linewidth,
    height=0.9\textheight,
    keepaspectratio
  ]{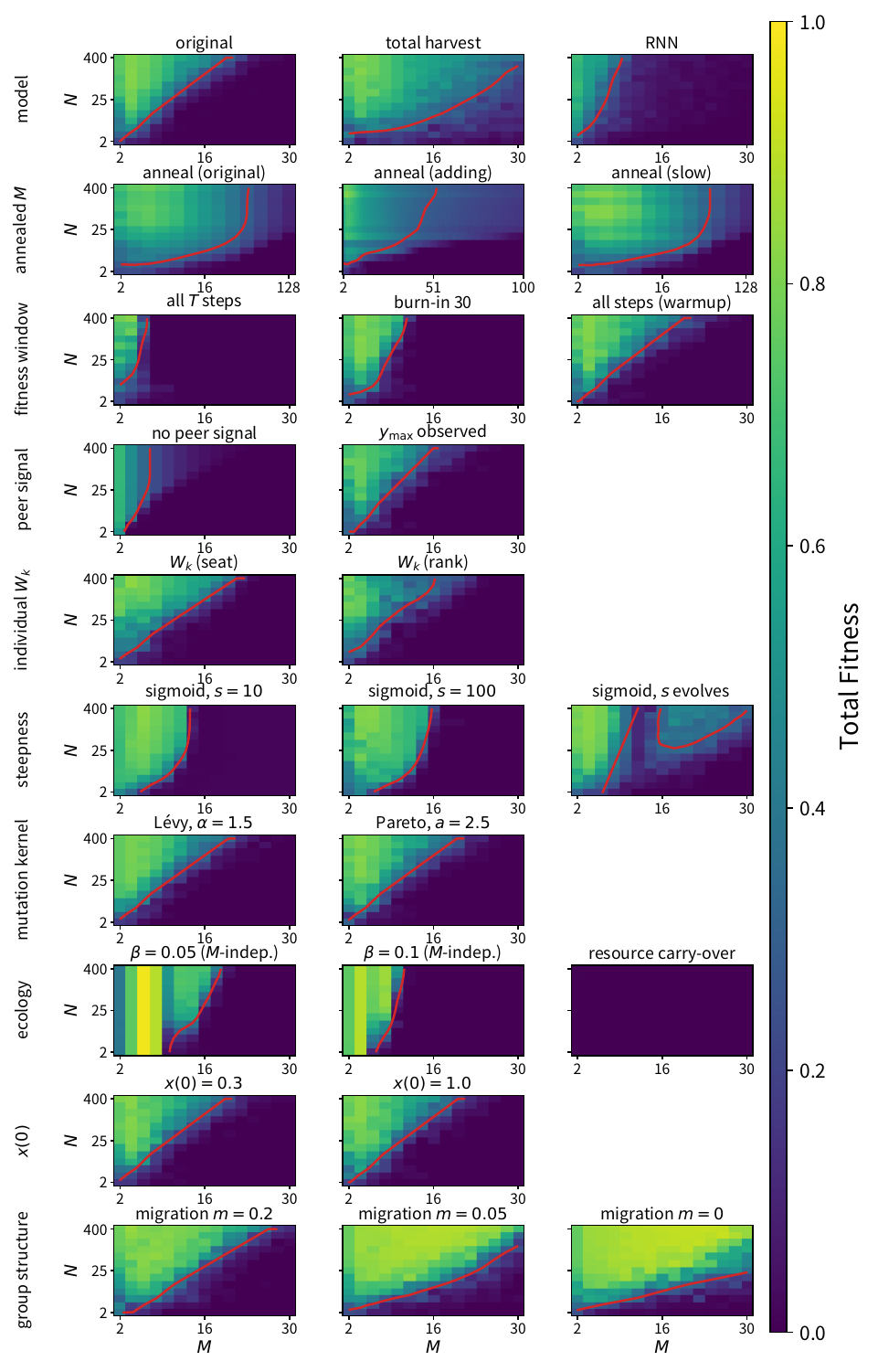}

  \caption{Robustness of sustainability across alternative model specifications.
  }
  \label{fig:DS_game_Mplayer_scaling}
\end{center}

\begin{figure}[!htbp]
  \centering
   \includegraphics[width=\linewidth]{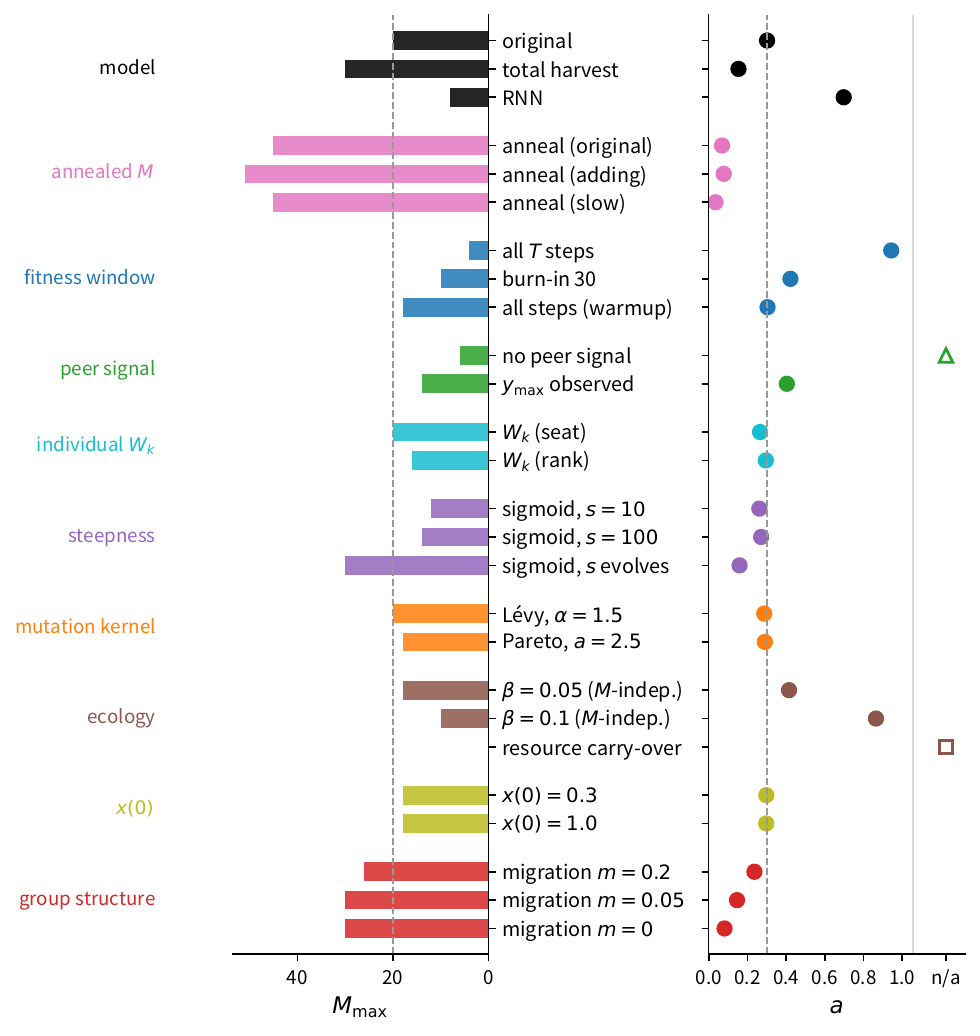}
   \caption{Summary of sustainability across alternative model specifications. The model variants correspond to those in Fig.~\ref{fig:DS_game_Mplayer_scaling} and are grouped by colour according to the component varied. (Left) $M_{\max}$ is the largest group size attaining mean total fitness $>0.3$ at $N=400$. Larger values indicate that sustainability evolves at larger group sizes. (Right) $a$ is the exponential rate obtained by fitting $N_c\propto \exp(aM)$, where $N_c$ is the minimum number of groups required for total fitness to exceed $0.3$. Smaller $a$ indicates a slower increase in the number of groups required as $M$ grows, which indicates smaller management burden. Circles show the estimated $a$. Conditions to the right of the light grey solid line (n/a) are those for which $a$ could not be estimated: the open triangle indicates that sustainable cells occurred but too few interior crossings were available, and the open square indicates that sustainability was not observed anywhere in the sampled range. Grey dashed lines show the original-model values of $M_{\max}$ and $a$. Sample sizes are $100$ trials per cell for the original model and $30$ for all other conditions.}
\label{fig:DS_game_Mplayer_scaling_exponent}
\end{figure}

\begingroup
\small

\noindent
    Each heatmap shows mean total fitness as a function of group size $M$
    and number of groups $N$. Colours indicate total fitness and red curves
    mark the boundary of sustainability (total fitness $>0.3$), whose exponents are shown in Fig. \ref{fig:DS_game_Mplayer_scaling_exponent}.
    \emph{Model}: ``original'' denotes the baseline model; ``total harvest'' uses the total harvest as a fitness, while the plotted total fitness remains wealth based; and ``RNN'' replaces the threshold rule with a recurrent neural network with four hidden units and inputs $x$, $My_i$, and $M\ev{y}$. \emph{Annealed $M$}: group size is increased either by multiplying $M$ by $\sqrt{2}$ every $1000$ generations [``anneal (original)''], adding one player every $500$ generations [``anneal (adding)''], or multiplying $M$ by $\sqrt{2}$ every $3000$ generations [``anneal (slow)'']. \emph{Fitness window}: fitness is calculated over all $T$ steps from the outset, after excluding the first $30$ steps, or over all steps after a $100$-generation warmup during which the first $100$ steps are excluded.
\emph{Peer signal}: players either receive no peer information, so decisions depend only on $x$ and $y_i$, or observe group-maximum wealth $y_{\max}$ instead of $\ev{y}$. \emph{Individual $W_k$}: $M$ evolving weights corresponding to every group member, with other players indexed either by seat position or by descending wealth rank.
\emph{Steepness}: harvesting occurs with probability $\sigma(kz_i) = 1 / (1+\exp(-kz_i))$, where $z_i=x+S_iMy_i+A_iM\ev{y}$ and the scale-independent steepness $s=k\sqrt{S_i^2+A_i^2}$ is fixed at $10$ or $100$, or evolves. With $s$ fixed, the total fitness exceeds that of the original model in almost every cell
where sustainable resource use evolves with $M \geq 6$, while the range of $M$ is narrower
(see Fig.~\ref{fig:DS_game_Mplayer_scaling_exponent}).
This gain corresponds to the positive effect of monitoring noise in Fig.~\ref{fig:DS_game_Mplayer_noise}:
a steepness of $s = 10$ and $100$ is equivalent to monitoring noise
$\sigma_p \approx 0.3$ and $0.03$, respectively.
\emph{Mutation kernel}: mutational steps of the evolving parameters are drawn from a
heavy-tailed distribution instead of the baseline Gaussian, either a symmetric
L\'evy $\alpha$-stable distribution with stability index $\alpha=1.5$ or a Pareto
distribution with exponent $a=2.5$ and a random sign, so that rare large jumps
occur with non-negligible probability.
\emph{Ecology}: the harvesting fraction $\beta$ is fixed at $0.05$ or $0.1$, rather than scaled as $0.9/M$, or the final resource level of one generation is carried over to the next instead of being reset.
\emph{$x(0)$}: the resource level at the start of each generation is initialised at
$x(0)=0.3$ or $x(0)=1.0$ instead of the baseline $x(0)=0.1$.
\emph{Group structure}: offspring remain associated with their parental group. Each position in a descendant group is filled from its founding-group pool with probability $1-m$ and from the global offspring pool with probability $m$, so smaller $m$ preserves stronger group structure. The original model represents $m = 1$. Each cell is averaged over $100$ trials for the original model and $30$ for all other conditions.
\par
\endgroup

\clearpage

\end{document}